\documentclass[printer]{aa}

\usepackage{amsmath}
\usepackage{amssymb}
\usepackage{soul}
\usepackage{natbib}
\usepackage{graphicx}
\usepackage{enumerate}
\usepackage{datetime}
\usepackage{color}
\usepackage{hyperref}
\usepackage[flushleft]{threeparttable}
\usepackage{booktabs}
\usepackage{tabulary}
\usepackage{multirow}
\usepackage{subcaption}
\usepackage{draftcopy}
\usepackage{dblfloatfix}

\newcommand{\gRad}{g_\mathrm{rad}}
\newcommand{\vturb}{v_\mathrm{turb}}
\newcommand{\vturbObs}{v_\mathrm{turb,\,Obs}}
\newcommand{\vturbTh}{v_\mathrm{turb,\,Theory}}
\newcommand{\cs}{c_\mathrm{s}}
\newcommand{\cseff}{c_\mathrm{s,eff}}
\newcommand{\Rpmod}{R_\mathrm{p,mod}}
\newcommand{\rhoo}{\rho_\ast}

\newcommand{\Msun}{M_\odot}
\newcommand{\Rsun}{R_\odot}
\newcommand{\Lsun}{L_\odot}
\newcommand{\Teff}{T_\mathrm{eff}}

\newcommand{\ddr}[1]{\frac{\partial #1}{\partial r}}

\newcommand{\code}[1]{\texttt{#1}}

\newcommand\Tstrut{\rule{0pt}{2.6ex}}         % = `top' strut
\newcommand\Bstrut{\rule[-0.9ex]{0pt}{0pt}}   % = `bottom' strut

\definecolor{edit}{rgb}{0.5,0.0,0.13}

\definecolor{dylan}{rgb}{0.78, 0.08, 0.52}

\definecolor{jon}{rgb}{0.13, 0.67, 0.8}

\definecolor{leen}{rgb}{0.4, 0.61, 0.0}

\definecolor{alex}{rgb}{0.57, 0.36, 0.51}

\begin{document}

\title{Analytic, dust-independent mass-loss rates for red supergiant winds initiated by turbulent pressure}
\titlerunning{Analytic RSG mass-loss rates from turbulent pressure}

   \author{
          N. D. Kee\inst{1},
          J. O. Sundqvist\inst{1},
          L. Decin\inst{1},
          A. de Koter\inst{1,2},
          \and H. Sana\inst{1}
          }
          \authorrunning{N. D. Kee et al.}

	\institute{Institute of Astronomy, KU Leuven, Celestijnenlaan 200D, B-3001 Leuven, Belgium\\
     \email{dylan.kee@kuleuven.be}
\and
	Anton Pannekoek Institute, University of Amsterdam, Science Park 904, 1098XH Amsterdam, The Netherlands\\
             }

   \date{}

  \abstract
  % context heading (optional), leave it empty if necessary  
   {Red supergiants are observed to undergo vigorous mass-loss.
   However, to date, no theoretical model has succeeded in explaining the origins of these objects' winds. 
   This strongly limits our understanding of red supergiant evolution and Type II-P and II-L supernova progenitor properties.
   }
   {We examine the role that vigorous atmospheric turbulence may play in initiating and determining the mass-loss rates of red supergiant stars.}
   {We analytically and numerically solve the equations of conservation of mass and momentum, which we later couple to an atmospheric temperature structure, to obtain theoretically motivated mass-loss rates.
   We then compare these to state-of-the-art empirical mass-loss rate scaling formulae as well as observationally inferred mass-loss rates of red supergiants.}
   {We find that the pressure due to the characteristic turbulent velocities inferred for red supergiants is sufficient to explain the mass-loss rates of these objects in the absence of the normally employed opacity from circumstellar dust.
   Motivated by this initial success, we provide a first theoretical and fully analytic mass-loss rate prescription for red supergiants. We conclude by highlighting some intriguing possible implications of these rates for future studies of stellar evolution, especially in light of the lack of a direct dependence on metallicity.}
   {}


   \maketitle

\section{Introduction}\label{sec:intro}

The lack of a satisfactory theory explaining the strong, $> 10^{-7}$ $\rm \Msun ~yr^{-1}$, mass loss for evolved massive stars on the red supergiant branch has been a long-standing problem in our understanding of these objects \citep[see][for a recent discussion of this and other current puzzles in studying red supergiants]{Lev17}.
While this considerable gap in our knowledge has been patched over somewhat by 
empirical rates and scaling formulae \citep[see, e.g.,][for a review]{MauJos11}, the overall disagreement between any given two of these leaves those attempting to model red supergiants directly, or attempting to model stellar evolution including or approaching this phase, with a rather untenable problem.
Namely, one must somehow distinguish and select between rates that may differ from each other by as much as four orders of magnitude in some parts of the parameter space
without a general understanding of the underlying process that determines this mass loss.

In order to understand why theories generally struggle so much in this region, it is important to highlight some general features of previous modeling attempts.
In analogy to the lower-mass asymptotic giant branch (AGB) stars, it has generally been assumed that the primary driving mechanism of red supergiant winds is 
radiation pressure on dust grains \citep[see, e.g.,][for a review]{HofOlo18}.
For AGB stars, pulsations provide an atmospheric piston levitating material off the stellar surface and up to a region where the temperature has dropped far enough 
to allow dust to condense \citep[e.g.,][]{BlaHof12,BlaHof13}.
However, as dust nucleation is also a density-dependent process, it is essential not only that material reaches this region but that enough material is levitated to make the dust formation efficient.
Due to both the lower pulsational amplitudes inferred for red supergiants \citep[e.g.,][]{WooBes83} and their higher effective temperatures relative to AGB stars, applying similar models to red supergiants thus requires 
gas to reach a greater height in comparison to the stellar radius, while less material is expected to be levitated in the first place.
This implies a strongly decreased efficiency of dust condensation.
Modeling attempts have been generally unsuccessful in generating the atmospheric extensions of red supergiants necessary to put enough material at the dust sublimation front to recover the observed mass-loss rates of red supergiants \citep[e.g.,][]{ArrWit15}.

An alternative suggestion has been that pulsational motions might be accompanied or replaced by significant atmospheric turbulence \citep[e.g.,][]{GusPle92,JosPle07}, and that this turbulence might be seeded by the vigorous convection expected in the atmospheres of red supergiants \citep[see][for a theoretical treatment of convection in red supergiant envelopes]{FreSte12}.
At the moment, these convection simulations produce only modestly extended atmospheres compared to what is inferred from observations \citep[e.g.,][]{ArrWit15}.
Nevertheless, observations of red supergiants do suggest that the outer layers of these stars are indeed very turbulent \citep[e.g.,][]{JosPle07,OhnWei17}.

Therefore, inspired by the work of \cite{GusPle92} and \cite{JosPle07}, we here undertake the derivation of an analytic, theoretical mass-loss rate that focuses on these observed turbulent velocities present in the atmospheres of red supergiants.
Doing so, we find that the inferred turbulent motions are, alone, sufficient to explain the mass-loss rates of red supergiants even in absence of any dust opacity.
This model then can be seen as a turbulent pressure driven extension of the classical thermal pressure driven Parker wind \citep{Par58}.
In contrast to the solar wind and other applications of this theory \citep[for instance the ``warm wind'' model,][]{Hea75,RogLam75}, however, the total sound and turbulent speed can be much lower due to the lower surface gravity of red supergiants.
In Sect. \ref{sec:mdot_iso} we lay out the basic formalism for this model in an isothermal set-up, under which the mass-loss rate is fully analytic.
By imposing a temperature structure, we then extend the model in Sect. \ref{sec:mdot_noniso} by iteratively solving the relevent equations to converge to non-isothermal mass-loss rates.
We further fit over the difference between the isothermal and non-isothermal mass-loss rates to provide a theoretically motivated mass loss rate as an analytic function of stellar parameters and turbulent velocity.
In Sect.~\ref{sec:comp_w_emp_and_obs} we review some of the currently used empirical mass-loss rate prescriptions for red supergiants and compare our theoretical rate to these, as well as to the sample of observed red supergiants from \cite{JosPle07} and the observations of Antares by \cite{OhnWei17}.
Finally, we discuss some future directions for this model in Sect. \ref{sec:outlook}.

\section{ Analytic mass-loss rate from levitation of an isothermal atmosphere}\label{sec:mdot_iso}

\subsection{Derivation of the mass-loss rate}\label{sec:mdot_derivation}

To begin, we first consider the relevant equations that must be satisfied at all points in an isothermal flow, namely conservation of mass and momentum.
For a purely radial, spherically symmetric outflow in a steady state these are 

\begin{align}
    &v\ddr{\rho} + \rho \ddr{v} + \frac{2 \rho v}{r} = 0 \label{eqn:cons_mass_iso}\\
    &v \ddr{v} = -\frac{1}{\rho}\ddr{P} - \frac{G M_\ast}{r^2} \label{eqn:cons_mom_iso}\;,
\end{align}
expressed in terms of radial velocity $v$, density $\rho$, and total pressure $P$ as functions of spherical radius $r$, as well as stellar mass $M_\ast$ and gravitational constant $G$.
%\dylan{etc? or all symbols have standard meaning etc.}

%}
%Assuming that this radial component of flux only falls off like luminosity $L$ over radius squared, i.e. $F_r = L/(4 \pi r^2)$,
%
%\begin{equation}
%    g_\mathrm{rad} = %\frac{\kappa L}{4 \pi r^2 c} %= \Gamma \frac{G %M_\ast}{r^2}
%\end{equation}
%From both observational and theoretical studies of red supergiants, we expect turbulent pressure to play an important role in describing the structure of the stellar atmosphere and wind \dylan{cite model atm. and obs. inferred vturb vs. cs}
%Therefore, a

As is done in standard static model atmosphere computations (see, e.g., \citealt{GusEdv08}; their Sect.~3) %\citep[see, e.g.][their Sect. 3]{GusEdv08}
pressure is expressed as the sum of thermal pressure $P_\mathrm{therm}$~=~$\rho c_\mathrm{s}^2$, turbulent pressure $P_\mathrm{turb}=\rho v_\mathrm{turb}^2$, and radiation pressure $P_\mathrm{rad}$ where the characteristic velocities of the first two are respectively the sound speed $\cs$ and the turbulent speed $\vturb$.
As concerns the radial component of radiation pressure gradient, we associate this with the acceleration $g_\mathrm{rad}$ such that

\begin{equation}\label{eqn:g_rad}
    -\frac{1}{\rho}\ddr{P_\mathrm{rad}} = g_\mathrm{rad} = \frac{\kappa F_{r}}{c} = 
    \frac{\kappa L_\ast}{4 \pi r^2 c} = \Gamma \frac{G M_\ast}{r^2}\;,
\end{equation}
%
%\jon{
with $c$ the speed of light, $\kappa$ the flux weighted mean opacity $[\rm cm^2/g]$, and $F_r$ the radial component of the flux, related to the stellar luminosity $L_\ast$ as
$F_r = L_\ast/(4 \pi r^2)$.
The final equality further introduces the Eddington factor $\Gamma \equiv \kappa L_\ast/(4 \pi G M_\ast c)$.
%, both for the moment taken as constants.
Explicitly replacing pressure with this combination and substituting Eq. \ref{eqn:cons_mass_iso} into Eq. \ref{eqn:cons_mom_iso} to eliminate density gradient terms, we find

\begin{equation}\label{eqn:cons_mass+mom_iso}
%    \begin{split}
        v\left(1-\frac{\cs^2 +\vturb^2}{v^2}\right) \ddr{v} =
        \frac{2 \left(\cs^2 +\vturb^2\right)}{r}  - \frac{G M_\ast \left(1-\Gamma\right)}{r^2}\;,
%    \end{split}
\end{equation}
where we have assumed that $\vturb$ is constant. We discuss this approximation further below.

We note that Eq. \ref{eqn:cons_mass+mom_iso} is the same equation as is solved for an isothermal, pressure driven Parker wind \citep{Par58}, only now with a modified ``effective sound speed'', $\cseff \equiv \sqrt{\cs^2+\vturb^2}$.
At the location where the flow velocity reaches this effective sound speed the left side of Eq. \ref{eqn:cons_mass+mom_iso} goes to zero, and therefore the right side must also go to zero.
Solving 
%the right side of the equation 
for this criteria yields that the location of this critical point in the wind outflow is given by a modified Parker radius,

\begin{equation}\label{eqn:Rpmod}
    \Rpmod = \frac{G\, M_\ast\, (1-\Gamma)}{2\, (\cs^2+\vturb^2)}\;.
\end{equation}
This then 
%implies 
suggests that the problem of wind acceleration here can be envisioned 
to consist of two parts.
First, a low-speed near ``levitation'' of material to this modified Parker radius at which $v = \cseff$, and, second, a subsequent acceleration of the material to infinity.
%in two parts, namely some low-speed nearly ``levitation'' of material to this modified Parker radius at which $v = \cseff$, and then a subsequent acceleration of the material to infinity.

Below the modified Parker radius the contribution of the advection term ($v~\partial v/\partial r$) in Eq. \ref{eqn:cons_mom_iso} is almost negligible so that the equation reduces to the standard equation for hydrostatic equilibrium, 

\begin{equation}\label{eqn:hydrostatic_density}
    \frac{1}{\rho}\ddr{\rho} = - \frac{G\, M_\ast\, (1-\Gamma)}{(\cs^2 + \vturb^2)\,r^2}\;.
\end{equation}

%\jon{
Typically, the Eddington factor and, as alluded to above, the turbulent speed may be expected to vary in radius. 
However, under the assumption that 
%$\Gamma$ and $\vturb$ are nearly constant in radius, 
%this variation is only mild, we may replace the exact expressions $\Gamma(r)$ and $\vturb(r)$ with some representative constant values $\Gamma$ and $\vturb$, which then in general would represent some appropriate spatial means 
this variation is only mild, we may replace the exact expression $\Gamma(r)$ with $\Gamma$ and, as already done in Eq. \ref{eqn:cons_mass+mom_iso}, $\vturb(r)$ with $\vturb$. These representative constant values of $\Gamma$ and $\vturb$ then in general would designate some appropriate spatial means
(see also the discussion in \citealt{GusEdv08}). Then
%}
Eq. \ref{eqn:hydrostatic_density} can be integrated analytically from the stellar radius $R_\ast$ to an arbitrary radius $r \leq \Rpmod$ using the boundary value definition $\rho_\ast \equiv \rho(R_\ast)$ to give
%This has the analytic solution

\begin{equation}\label{eqn:rhor}
    \rho(r) = \rhoo \exp\left[-\frac{R_\ast}{H}\left(1 - \frac{R_\ast}{r}\right)\right]\;,
\end{equation}
where we have made the substitution

\begin{equation}\label{eqn:scale_height}
    \frac{H}{R_\ast} \equiv \frac{R_\ast\,(\cs^2 + \vturb^2)}{G\, M_\ast\, (1-\Gamma)} = 2\,\frac{\cs^2 + \vturb^2}{v_\mathrm{esc,eff}^2} = \frac{1}{2}\,\frac{R_\ast}{\Rpmod}\;.
\end{equation}
Here $v_\mathrm{esc,eff}\equiv \sqrt{G M_\ast (1-\Gamma)/R_\ast}$ denotes the escape speed from $R_\ast$ with $\sqrt{1-\Gamma}$ accounting for the reduction in effective gravity due to radiative acceleration.

In order to determine the value of $\rhoo$, we make use of the definition of the stellar radius in optical depth space, $\tau(R_\ast) \equiv 2/3$, from a spherical extension of the classical planar gray model atmosphere 
%\citep[][and see further Sect. \ref{sec:mdot_noniso}]{Luc71}. 
(\citealt{Luc71}; and see further Sect. \ref{sec:mdot_noniso}).
%No, that should be ok for now. \dylan{cite \cite{SanPul97}?}
%Under such a spherical extension of the grey model atmosphere, 
Within this model, one defines a "spherically modified" optical depth scale as

\begin{equation}\label{eqn:sph_mod_tau}
    \tau(r) \equiv \int_r^\infty \kappa\, \rho\, \left(\frac{R_\ast}{r}\right)^2 dr\;, 
\end{equation}
and, as such,

\begin{equation}\label{eqn:sph_mod_tau_Rstar}
    \kappa \int_{R_\ast}^\infty \rho(r) \left(\frac{R_\ast}{r}\right)^2 dr = \frac{2}{3}\;,
\end{equation}
where we have again used the above approximation that $\Gamma$, and therefore $\kappa$, is independent of radius.
Inserting the hydrostatic density profile from Eq. \ref{eqn:rhor} into Eq. \ref{eqn:sph_mod_tau} yields an analytic value of $\rhoo$,

\begin{equation}\label{eqn:rhostar}
    \rhoo = \frac{2}{3\,\kappa\, H}\left[1 - \exp\left(-\frac{R_\ast}{H}\right)\right]^{-1} \;.
\end{equation}

Before combining the above terms to derive a turbulent pressure driven mass-loss rate, it is important to examine a few of the assumptions that went into the derivation.
First, we consider the approximation that the advection term is negligible for $r < \Rpmod$.
If we relax this approximation, we find that integrating Eq. \ref{eqn:cons_mom_iso} from $R_\ast$ to an arbitrary radius $r$ gives

\begin{equation}
    \log\left(\frac{\rho(r)}{\rhoo}\right) = -\frac{R_\ast}{H}\left(1-\frac{R_\ast}{r}\right) -\frac{1}{2\,(\cs^2+\vturb^2)}\int_{R_\ast}^r \ddr{v^2} dr\;.
\end{equation}
While in general solving this equation at an arbitrary radius requires numerical integration, the integration from $R_\ast$ to $\Rpmod$ remains analytic as $v(R_\ast)^2/(\cs^2 + \vturb^2) \approx 0$ and $v(\Rpmod)^2/(\cs^2 + \vturb^2) \equiv 1$.
Comparing $\rho(\Rpmod)$ computed with and without the advection term shows that including this term reduces the density at the modified Parker radius by a constant factor $\exp(-1/2)$.
As this reduction is analytic and constant, we therefore include it in all calculations for the remainder of this section. 
%\jon{something seems to be missing here? (I hope I didn't remove anything..} 

\begin{figure}[h]
    \centering
    \includegraphics[width=0.48\textwidth]{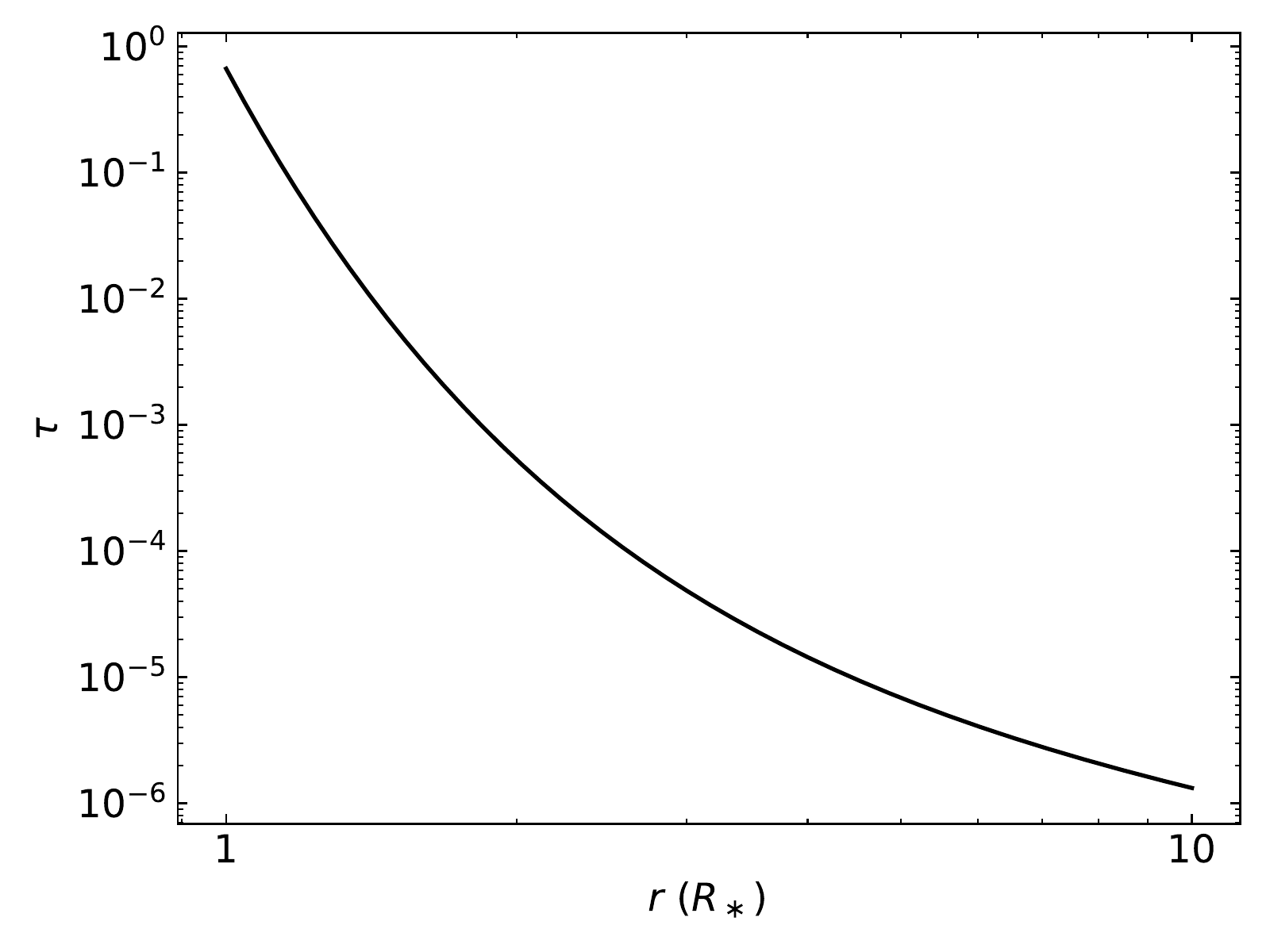}
    \caption{Optical depth calculated inward to r. Note that most of the optical depth is accumulated within a stellar radius of the star.}
    \label{fig:optical_depth}
\end{figure}

\begin{figure*}
    \centering
    \begin{subfigure}{0.32\textwidth}
        \includegraphics[width=\textwidth]{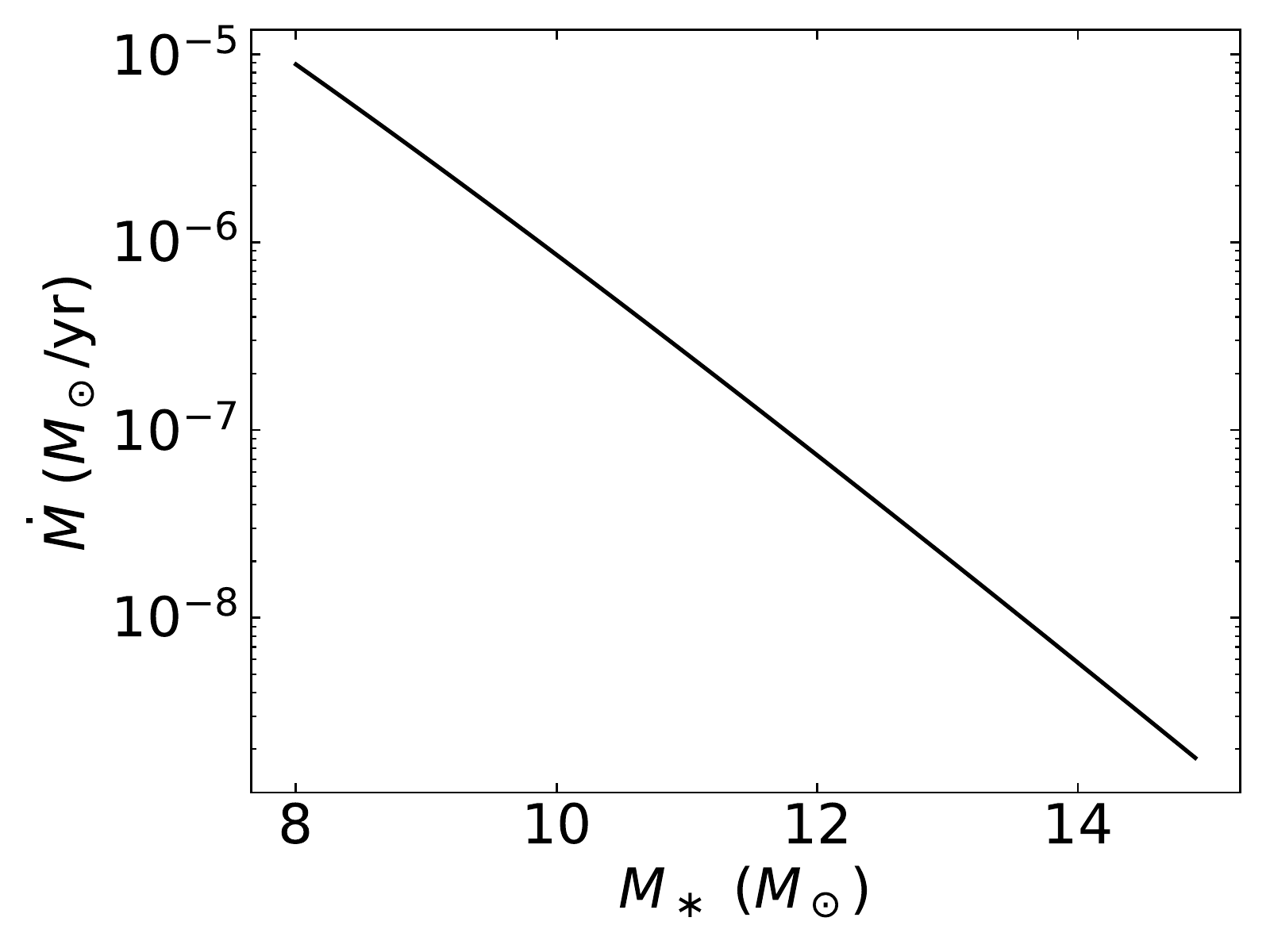}
    \end{subfigure}
    \begin{subfigure}{0.32\textwidth}
        \includegraphics[width=\textwidth]{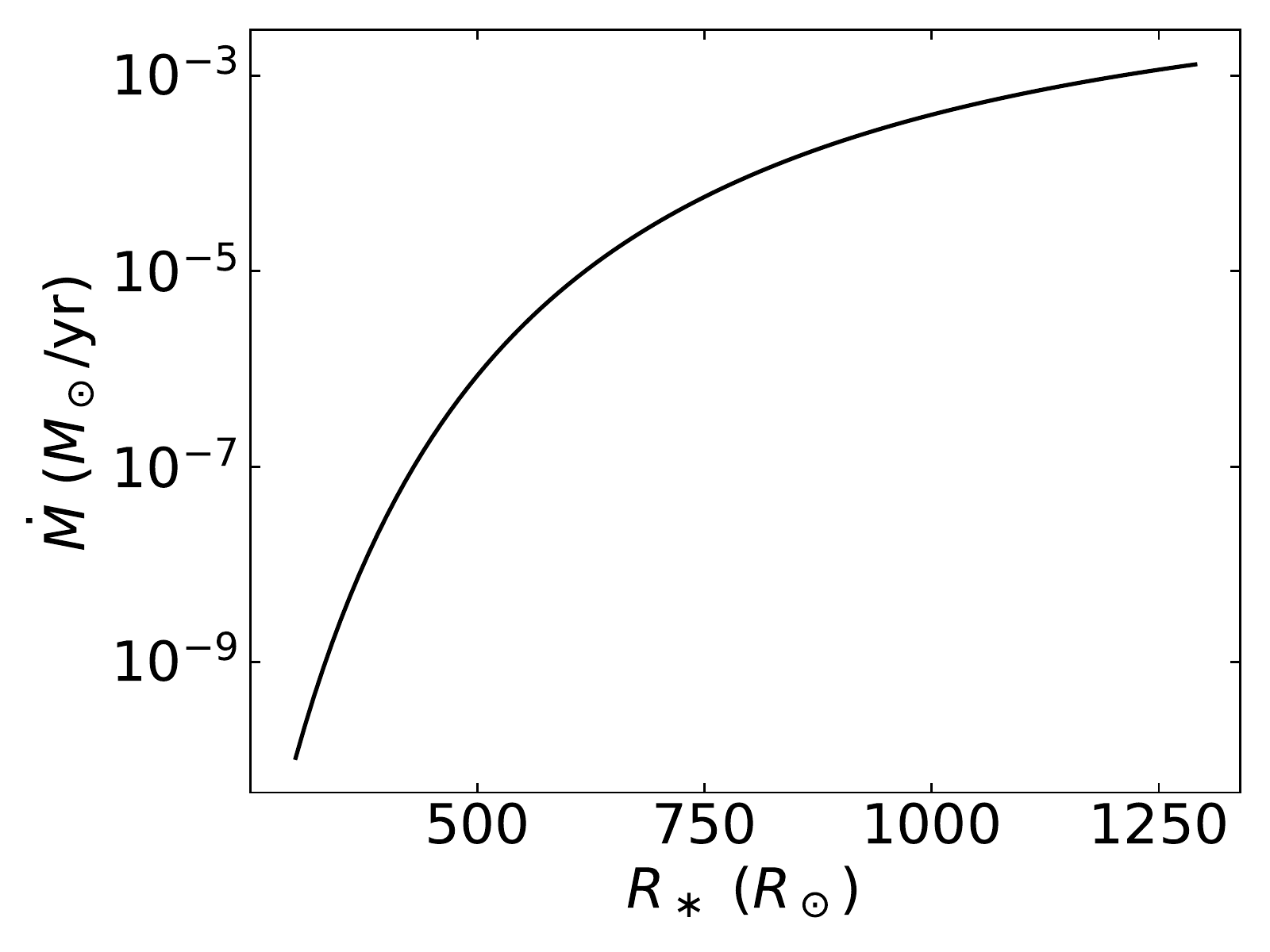}
    \end{subfigure}
    \begin{subfigure}{0.32\textwidth}
        \includegraphics[width=\textwidth]{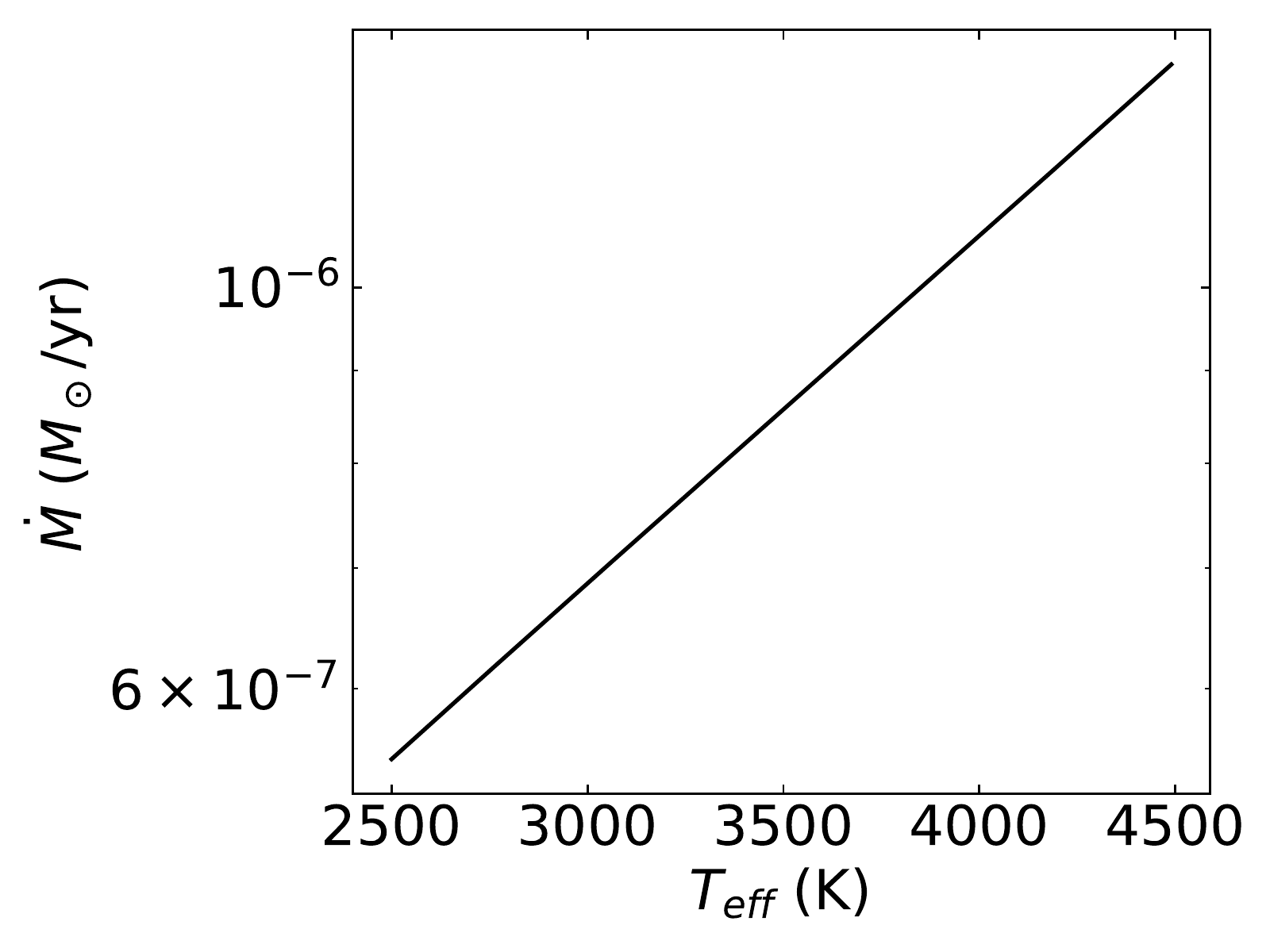}
    \end{subfigure}
    \caption{\label{fig:mdot_with_stellar} Dependence of the analytic mass-loss rate on stellar mass (left), stellar radius (center), and temperature (right).}
\end{figure*}

Related to this change in density, it is also important to examine whether the optical depth, and as such the base density $\rhoo$, is significantly impacted by the inclusion of the advection term or by a radial profile in the opacity.
To test this we plot $\tau(r/R_\ast)$ computed from the hydrostatic density structure with a constant opacity in Figure \ref{fig:optical_depth}, assuming that $H/R_\ast=0.07$, a characteristic value for a red supergiant atmosphere.
We note here, from examining Eqns. \ref{eqn:rhor}, \ref{eqn:sph_mod_tau}, and \ref{eqn:rhostar}, that $\tau(r/R_\ast)$ only depends on this chosen $H/R_\ast$ as the definition of $\tau(1) = 2/3$ results in $\kappa$ in Equations \ref{eqn:sph_mod_tau} and \ref{eqn:rhostar} cancelling.
As expected, almost all of the optical depth is accumulated within only a few scale heights of the stellar surface, here within about half a stellar radius of $R_\ast$.
This is well away from the modified Parker radius, here $\sim 7 R_\ast$, and therefore the optical depth used to define the stellar radius and $\rhoo$ is all accumulated in a region where the advection term is indeed negligible.
Similarly, this means that $\kappa$ and $(\vturb^2 + \cs^2)$ only need to be constant over this same small spatial extent in order for $\rhoo$ to be unaffected by any variations they may have.
Sect. \ref{sec:mdot_noniso} will revisit the implications of allowing $\cs$ to take on a realistic radial profile.

%\jon{Similar considerations hold regarding the above neglect of a potential spatial variation of $\vturb^2 + \cs^2$.} 

Finally, to turn the terms derived thus far into a mass-loss rate, we note that Eq. \ref{eqn:cons_mass_iso} is equivalent to the constraint $\rho v r^2 =$ const.;
%Using this we can define 
the total mass-loss rate is this conserved quantity integrated over all solid angle, $\dot{M} \equiv 4 \pi \rho v r^2$.
Further, as this quantity is independent of radius, we can simply take its value at the modified Parker radius

\begin{equation}\label{eqn:mdot_an}
    \dot{M} = 4\, \pi\, \rho(\Rpmod)\, \sqrt{\cs^2+\vturb^2}\, \Rpmod^2 \;,
\end{equation}
as the total mass-loss rate of the star.
Combining Eqs. \ref{eqn:rhor}, \ref{eqn:scale_height}, and \ref{eqn:rhostar} and accounting for the additional factor $\exp(-1/2)$ from advection gives $\rho(\Rpmod)$ to be

\begin{equation}\label{eqn:rhoRp}
\rho(\Rpmod) = \frac{4}{3} \frac{R_{\rm p,mod}}{\kappa R_\ast^2} \frac{\exp\left[-\frac{2 R_{\rm p,mod}}{R_\ast} +\frac{3}{2}\right]}{1-\exp\left[-\frac{2 R_{\rm p,mod}}{R_\ast}\right]}.
\end{equation}
All quantities in Eqs. \ref{eqn:mdot_an} and \ref{eqn:rhoRp} are analytically known within this formalism, such that these expressions together offer a fully analytic mass-loss rate.
%\jon{I would suggest to introduce a new subseciton for the below paragraph, or alt. move it to Sect. 2.2, since I think this last sentence would be a really neat "end" of this section...} 

%\subsection{Applicability of the model}\label{sec:applicability}

\subsection{Scaling of mass-loss rate with key parameters of the model}\label{sec:iso_scaling}

In understanding the regimes of stellar parameters where such a mass-loss rate model is and is not viable, it is important to note that $H/R_\ast = R_\ast/(2 \Rpmod)$ as shown in Eq. \ref{eqn:scale_height}.
This means that the same scale which controls the exponential stratification of density also controls where the critical point lies.
Therefore, $H/R_\ast$ must be relatively large compared to, for instance, the Sun in order to prevent the exponential density stratification of the atmosphere from driving $\dot{M}$ to zero.
At the same time, $\Rpmod$ and by extension $R_\ast$ must be large (in absolute units) simply due to the $\Rpmod^2$ dependence of $\dot{M}$.

%\dylan{Think about whether to mention $\vturb$ const. or $\Teff$ const. here -- jon: made just short one-sentence suggestion, if correct?}

%Moreover, while we have neglected the velocity gradient in the region below the modified Parker radius for the derivation of the density structure, accounting for this leads to a factor exactly $\exp(-1/2)$ reduction in $\rho(\Rpmod)/\rhoo$ \citep[see, e.g.][]{LamCas19}.
%Given that this reduction is constant, we include it in the analyses in the following subsections.

The mass-loss rate derived in Sect. \ref{sec:mdot_derivation} is a function of only a few key parameters of the physical set-up, namely stellar mass $M_\ast$, stellar radius $R_\ast$, sound speed $\cs$, turbulent velocity $\vturb$, and opacity $\kappa$. 
Specifically,
\begin{equation}\label{eqn:mdot_scaling}
    \dot{M} \propto \frac{1}{\kappa} \left(\frac{R_\ast}{H}\right)^3 \exp\left(-\frac{R_\ast}{H}\right)\, R_\ast\;,
\end{equation}
where $H/R_\ast$ itself scales as
\begin{equation}
    \frac{H}{R_\ast} \propto \frac{R_\ast\,\left(\cs^2 + \vturb^2\right)}{M_\ast\,\left(1-\Gamma\right)}\;.
\end{equation}
We can trade the sound speed dependence for a stellar parameter by assuming for the following discussion that the wind has $\cs = \sqrt{k_\mathrm{B}\, \Teff/m_\mathrm{H}}$ with Boltzmann constant $k_\mathrm{B}$, mean molecular mass equal to the mass of a hydrogen atom $m_\mathrm{H}$, and temperature equal to the stellar effective temperature $\Teff = (L/(4\, \pi\, R_\ast^2\, \sigma))^{1/4}$, where $\sigma$ is the Stefan-Boltzmann constant.
%\jon{You need to define Teff somewhere above, since this is not a well-defined quantity in spherical models, and I think what you really mean here is that you're using it as the characteristic temperature to compute the sound speed, or? Similarly, in the above I don't see anything about the mean molecular weight, which also goes into this.} 
To investigate the behavior of this analytic expression, we begin by calculating the mass-loss from a star with parameters selected to be consistent with a ``typical'' RSG, namely $M_\ast = 10 \Msun$, $R_\ast = 500 \Rsun$, $\Teff = 3500$ K ($\cs = 5.4$ km s$^{-1}$), $\vturb = 15$ km s$^{-1}$, and $\kappa = 0.01$ cm$^2$ g$^{-1}$.
Note that here opacity is estimated from the OPAL opacity tables \citep[][]{IglFor96} as a lower limit to the total opacity as will be discussed further below.
For these parameters, $\dot{M} = 8.5\times 10^{-7}\; \Msun$ yr$^{-1}$, in the range of plausible mass-loss rates for a red supergiant \citep[see, e.g.,][]{JosPle07}.

For the remaining discussion in this section, we vary individual model parameters while holding all others to the values provided immediately above to show how $\dot{M}$ reacts to such variations.
The scaling of this mass-loss rate with stellar parameters is relatively straightforward as shown by Figure \ref{fig:mdot_with_stellar}.
Specifically, increasing the stellar radius or effective temperature over a reasonable range for red supergiants increases the mass-loss rate because increasing either of these increases $H/R_\ast$.
This goes into the exponential term in Eq. \ref{eqn:mdot_scaling} which increases faster than the power-law terms fall off.
%$\rho(r)$, $\rho(\Rpmod)$ increases faster than the variation of the other quantities in Eq. \ref{eqn:mdot_an}.
Conversely, increasing $M_\ast$ has the opposite effect on $H/R_\ast$ and therefore increasing stellar mass decreases mass-loss rate.
Note that $\Teff$ only appears through the sound speed, and thus only in the combination $\cs^2 + \vturb^2$.
Therefore, it is unsurprising that varying this parameter has a much weaker effect than either stellar mass or radius, a fact we return to later in the discussion.

\begin{figure}[ht]
    \centering
    \includegraphics[width=0.48\textwidth]{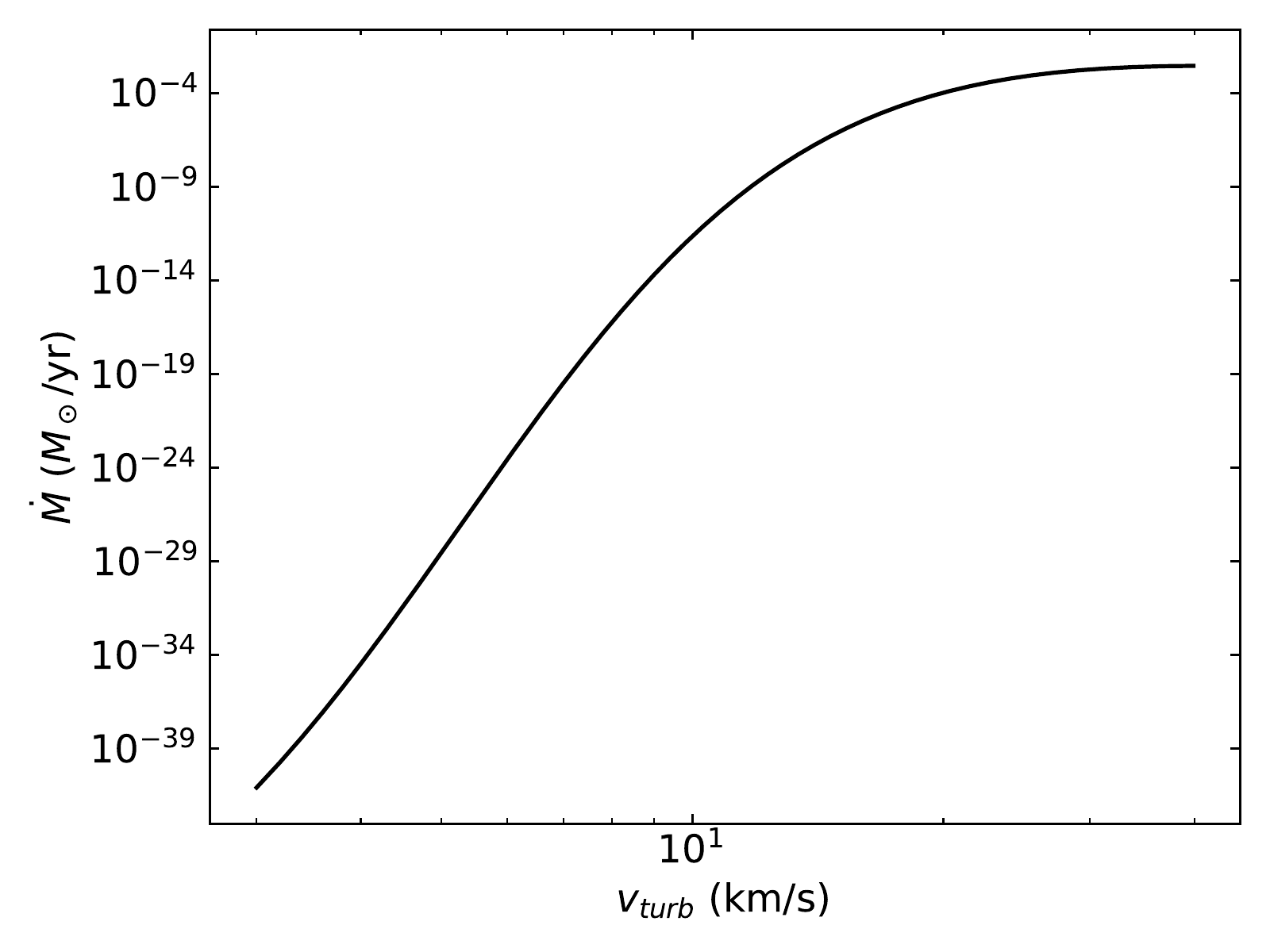}
    \caption{\label{fig:mdot_with_vturb} Dependence of the analytic mass-loss rate on turbulent velocity.}
\end{figure}

Next, we examine the scaling of mass-loss rate with $\vturb$, shown in Figure \ref{fig:mdot_with_vturb}.
In order to emphasize the importance of $\vturb$ as the driving mechanism for this model, we vary $\vturb$ from a minimum microturbulence for red supergiants ($\sim$ 3 km s$^{-1}$) up through the characteristic turbulent velocities inferred by \cite{JosPle07}.
As was the case with the variation of stellar parameters, for much of the range of $\vturb$ the dominant effect is on the scale height.
As $H/R \propto \vturb^2$ for $\vturb \gg \cs$, this dependence is quite steep.
However, note here that we allow a much larger range of variation in $\vturb$ than in the other parameters, which means that we see an additional regime that did not appear in the variation of stellar parameters, namely the asymptotic mass-loss rate reached at high $\vturb$.
Here $\exp(-R_\ast/H)$ begins to vary less rapidly with the increase in $H/R_\ast$ and the decrease of $\Rpmod$ and $\rhoo$ with increasing $H/R_\ast$ takes over.
The scaling is truncated at the point where increasing $\vturb$ means $\Rpmod < R_\ast$, as the model is no longer meaningful in this regime.
If $\vturb$ were allowed to vary to arbitrarily low values, a lower asymptote would also appear corresponding to the negligibly low mass loss that could be driven by $\cs$ alone.

\begin{figure*}
    \centering
    \begin{subfigure}{0.32\textwidth}
        \includegraphics[width=\textwidth]{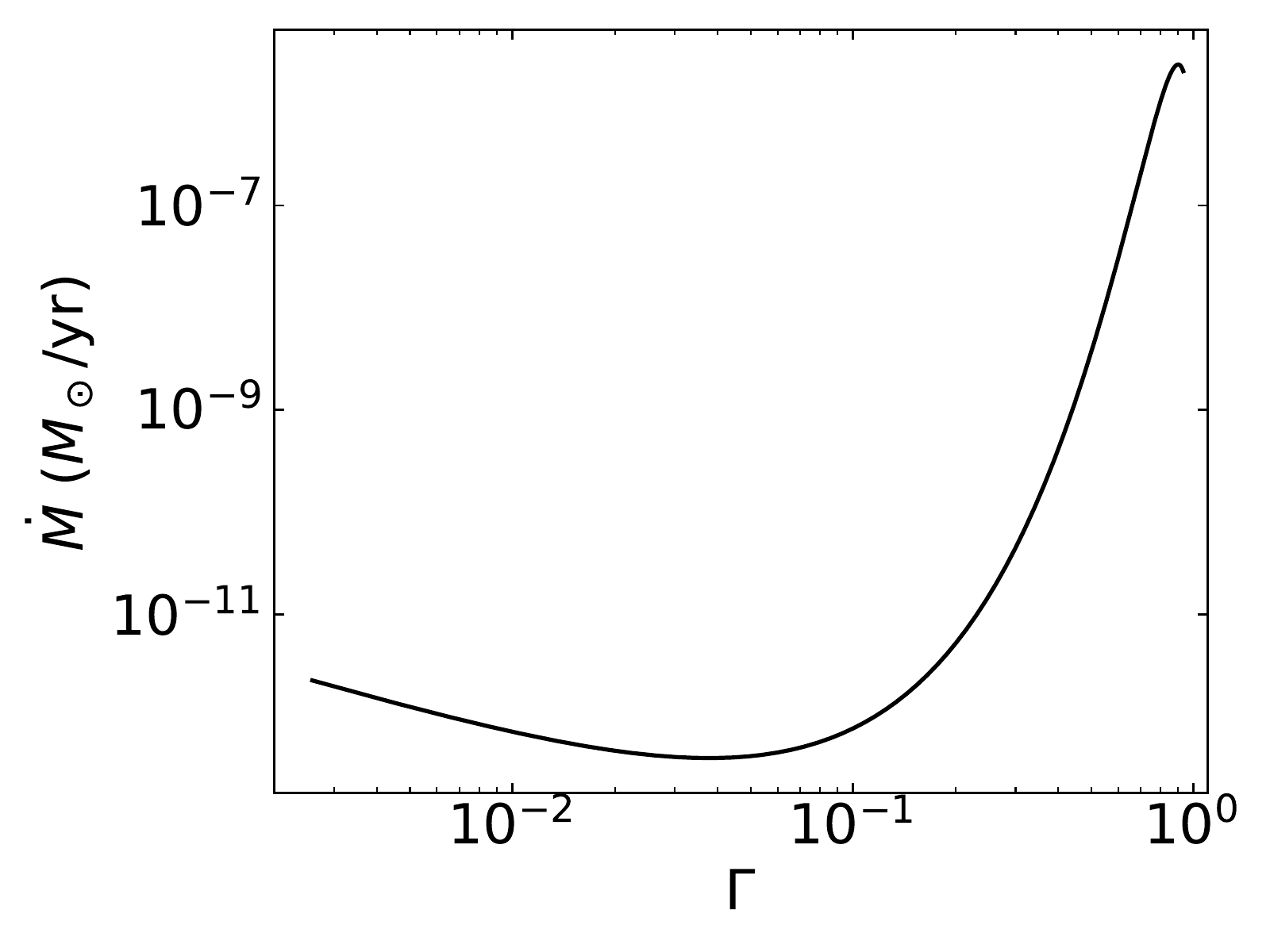}
    \end{subfigure}
    \begin{subfigure}{0.32\textwidth}
        \includegraphics[width=\textwidth]{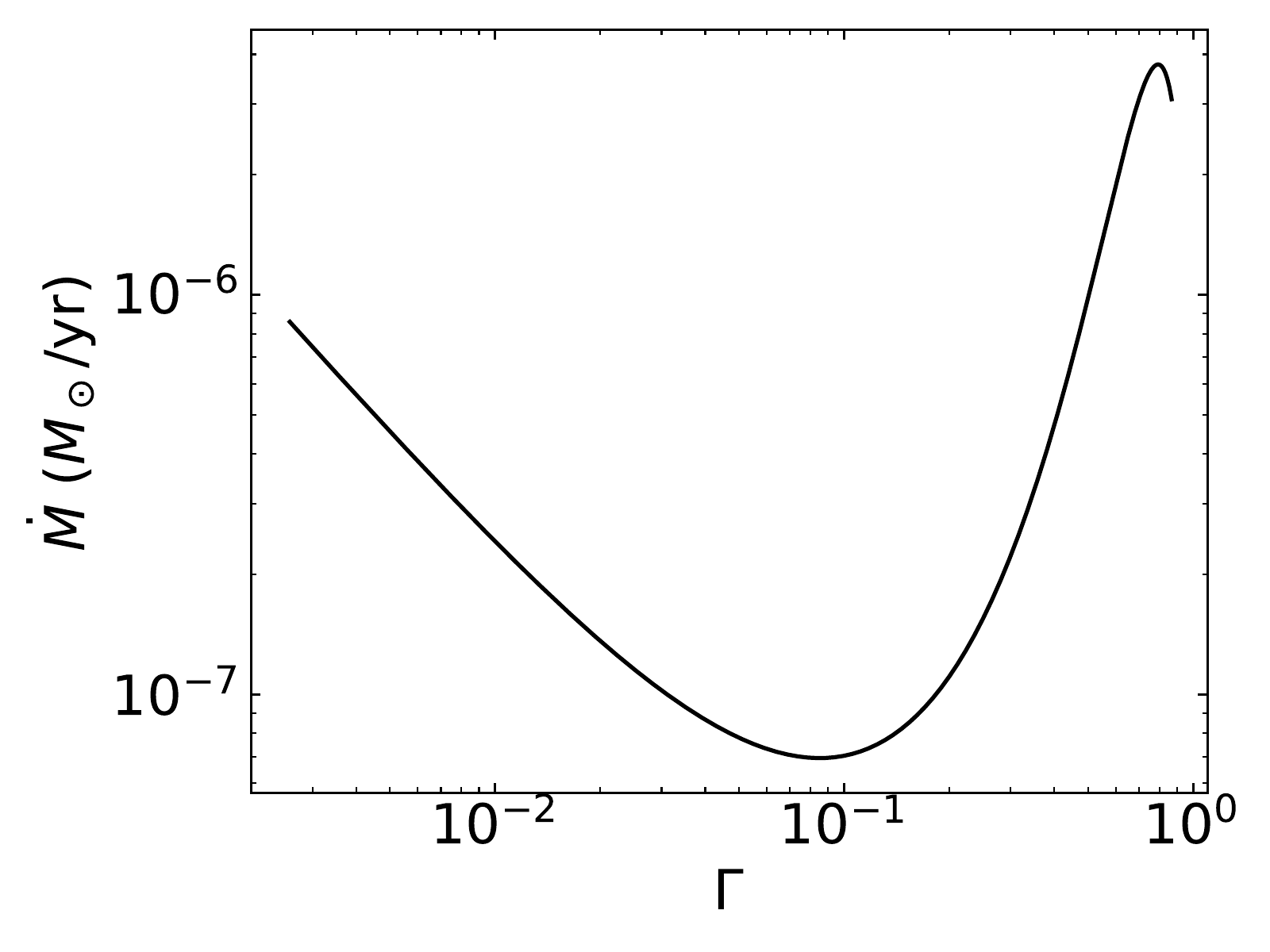}
    \end{subfigure}
        \begin{subfigure}{0.32\textwidth}
        \includegraphics[width=\textwidth]{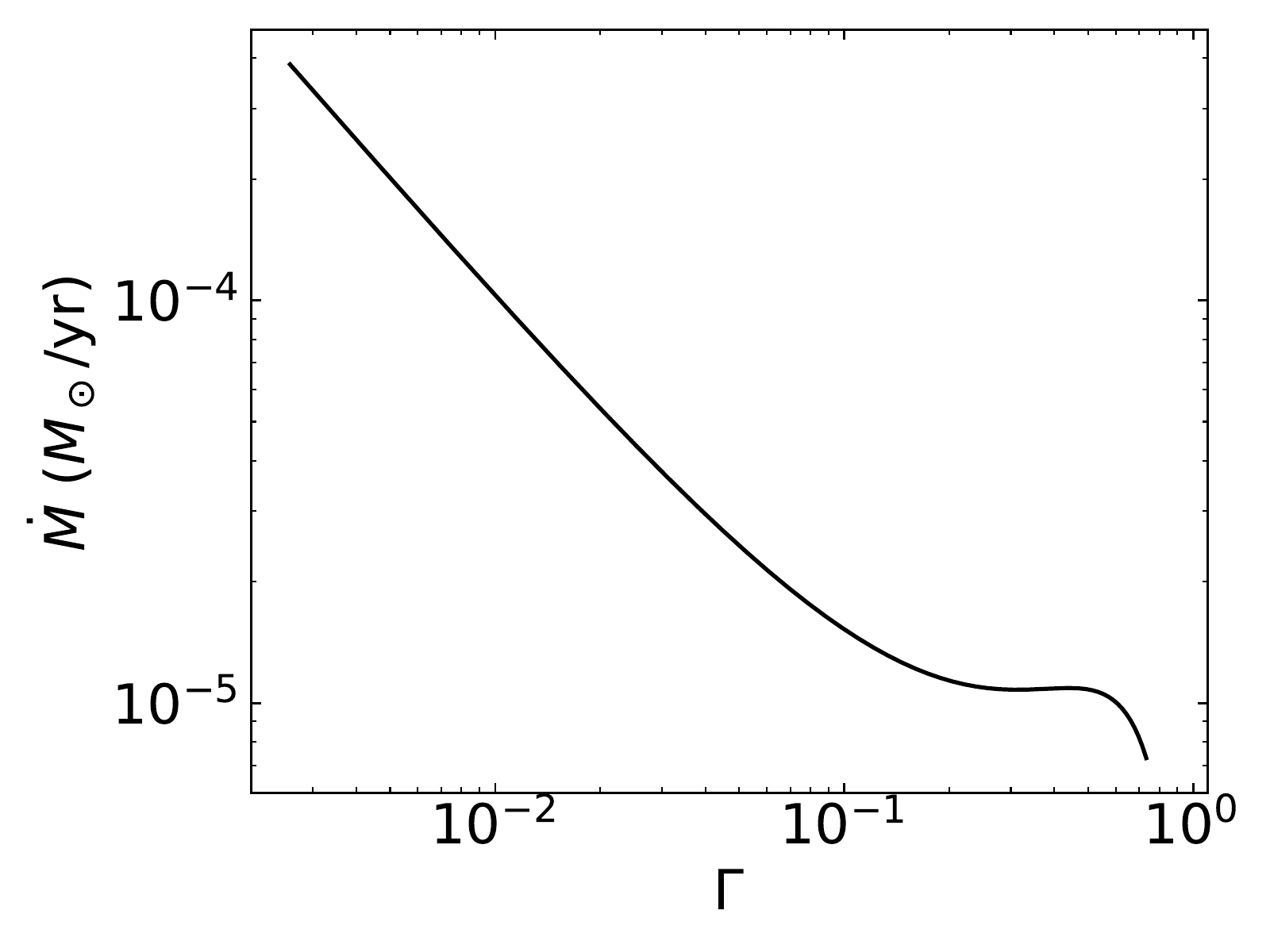}
    \end{subfigure}
    \caption{\label{fig:mdot_with_Gamma} Dependence of the analytic mass-loss rate on $\Gamma$ for $M_\ast = 10 \Msun$ and $\Teff = 3500$ K. The left panel uses $R_\ast = 500 \Rsun$ and $\vturb = 10$ km s$^{-1}$, the middle panel uses $R_\ast = 500 \Rsun$ and $\vturb = 15$ km s$^{-1}$, and the right panel uses $R_\ast = 1000 \Rsun$ and $\vturb = 15$ km s$^{-1}$.}
\end{figure*}

Finally, we examine the scaling of $\dot{M}$ with opacity, shown in Figure \ref{fig:mdot_with_Gamma}.
While the first calculation of $\dot{M}$ for characteristic red supergiant parameters above assumed the Rosseland mean opacity from OPAL, the acceleration of the wind means that a purely static opacity model may not be the most accurate one \citep[see also][]{GusPle92}.
In fact, the Doppler shifting of spectral lines from atomic and molecular species will increase the characteristic opacity scale of the wind, as discussed in Appendix \ref{app:line_force}.
In practice, the computation presented in Appendix \ref{app:line_force} for $\kappa$ and $\Gamma$ from these Doppler shifted spectral lines becomes computationally expensive to include in a model such as the one we discuss in the remainder of the paper.
Specifically, millions of spectral lines are involved \citep[see especially the tabulations of infrared spectral lines of water, e.g.,][]{BarTen06}, with the opacity of each line becoming position-dependent due both to the velocity gradient and the radial temperature variation.
In studies of hot, OB star winds, this is circumvented by introducing a distribution function that allows one to analytically approximate the cumulative effects of all these spectral lines.
As a first step, we have taken the brute force approach of computing the full sum but only for fixed hydrodynamic structures, thus allowing us to approximate the net effect of this additional radiative driving.
By computing the force from infrared spectral line lists of CO \citep[][]{Goo94}, TiO \citep[][]{McKMas19}, and H$_2$O \citep[][]{BarTen06} as a sample, we find that the net increase in flux-weighted mean opacity may be an order of magnitude or more over the basal OPAL opacity.
However, this preliminary study deserves further attention as it is based on analyses of fixed hydrodynamic structures, while line acceleration is a sensitive function of the velocity field.
Further, we have only included three molecules here, and also used a simplified radiative transfer treatment based on the so-called Sobolev approximation (see Appendix A), implying the real effect could be a bit different than the 1 dex mentioned above.
% implying the effect could be more than just 1 dex in $\Gamma$.}

Here we take a more general view of the effect of opacity variation, simply allowing $\Gamma$ to increase from its basal value as implied by the Rosseland mean opacity, $\Gamma_\mathrm{Ros}$.
As was the case for $\vturb$, we truncate the plot when $\Rpmod = R_\ast$ as this is again the limit at which the model breaks down.
To emphasize the complex behavior induced by varying $\Gamma$ and thus $\kappa$, we show the scaling of $\dot{M}$ for three models: $\vturb = 10$~km~s$^{-1}$ and $R_\ast = 500 \Rsun$ (left panel of Figure~\ref{fig:mdot_with_Gamma}), $\vturb = 15$~km~s$^{-1}$ and $R_\ast = 500 \Rsun$ (center panel of Figure~\ref{fig:mdot_with_Gamma}), $\vturb = 15$~km~s$^{-1}$ and $R_\ast = 1000 \Rsun$ (right panel of Figure~\ref{fig:mdot_with_Gamma}).
Note here that $\Gamma_\mathrm{Ros} = \kappa_\mathrm{R}\, L_\ast/(4\, \pi\, G\, M_\ast\, c) = \kappa_\mathrm{R}\, R_\ast^2\, \sigma\, \Teff^4/(G\, M_\ast\, c)$.
%\jon{seem to be missing some factors in teff above..} 
All parameters are the same for the three cases except $R_\ast$, such that this minimal point for $\Gamma$ is lower for the two models with $R_\ast = 500 \Rsun$ ($\Gamma_\mathrm{Ros} = 2.6\times10^{-3}$) than the one with $R_\ast = 1000 \Rsun$ ($\Gamma_\mathrm{Ros} = 1.0\times10^{-2}$).

In general, as increasing $\Gamma$ means increasing the radiation force available to 
%drive the wind, one might expect 
%\jon{
levitate the atmosphere,
%}
one might expect that $\dot{M}$ would monotonically increase with $\Gamma$.
However, it is important to recall that we choose our unique solution for the mass-loss rate by 
%setting $\tau=2/3$ 
%\jon{
requiring $\tau(R_\ast)=2/3$
%}
through Equation \ref{eqn:sph_mod_tau_Rstar}, which results in an inverse dependence of the stellar photosphere density on opacity in Equation \ref{eqn:rhostar}.
When $\Gamma$ is small and the wind is predominantly driven by turbulent pressure, this inverse dependence wins out and increasing $\Gamma$ and therefore $\kappa$ implies a reduction of the density of the wind and therefore also of the mass-loss rate to keep $R_\ast$ fixed.

As $\Gamma$ approaches unity, radiation then contributes more meaningfully to the total force budget of the wind and the expected increase manifests itself.
In the case that the turbulent pressure gradient is not already driving a strong wind mass loss (e.g. $\vturb$ and $R_\ast$ small as in the left panel of Figure \ref{fig:mdot_with_Gamma}) then radiative acceleration is able to drive a much stronger mass loss near $\Gamma$ unity than pressure could at low $\Gamma$.
In the case where turbulent pressure already drives a strong mass loss (i.e., $\vturb$ and $R_\ast$ large as in the right panel of Figure \ref{fig:mdot_with_Gamma}) the net effect is only to cancel out the reduction in $\dot{M}$ from the increased opacity.
However, even this increase (or flattening) is not able to hold out all the way to $\Gamma$ unity.
Eventually, as was the case with increasing $\vturb$, $\Rpmod$ is driven to $R_\ast$ and $\dot{M}$ drops off.
Again, this point merely denotes where such a turbulent pressure driven wind model breaks down.

%\jon{Tentatively moved this to here, since it deals with opacity (only for dust), so it seems it oculd fit below the discussion about general effects of such opacity. But I too am certainly open for moving it around further.} 
As mentioned in the introduction, it is intriguing that this model is able to recover reasonable mass-loss rates for red supergiants without appealing to dust opacity.
In fact, as $\Gamma_\mathrm{Ros}$ is small in the standard case we consider, this model does not appeal to significant radiative driving from any source.
%The general impression of 
One of the theories present in the literature is that red supergiant winds behave analagously to the winds of AGB stars with some atmospheric extension allowing for the condensation of dust, the opacity of which is the main mass-loss rate driver \citep[see e.g.][for a review]{HofOlo18}.
Instead, we here find that it is plausible that the extreme extension of their atmospheres by turbulent pressure alone could be the dominant part determining the red supergiant mass-loss process.

However, it is important to note here that opacity may still play an important role in altering the structure of the winds of red supergiants.
While we have here assumed a constant opacity as a function of distance from the star in order to examine the scaling of $\dot{M}$ with $\Gamma$, this is not expected to be the case.
As the wind cools away from the star, additional molecules and dust will form, thereby enhancing the wind opacity.
Even the continuum opacity is likely to shift as the Hydrogen anion is no longer the dominant continuum opacity source.
As discussed in the prior subsection, the effect of this enhanced opacity on at least the optical depth scale, and as such the density scale used in this model, may not be substantial as long as the opacity enhancement occurs beyond the first stellar radius or so.
This is because, as shown by Fig. \ref{fig:optical_depth}, the majority of optical depth is accumulated in a region quite near to the star making $\kappa$ in Eqn. \ref{eqn:rhostar} effectively an average over this near star region.
Enhancing the opacity further out would, however, contribute to extending the scale height of the atmosphere, and even potentially allow the wind to switch from being turbulence driven to radiation driven at some location.
Both these effects could increase $\dot{M}$ somewhat over what is presented here.
In light of these open items related to wind opacity, as we discuss further in Sect. \ref{sec:outlook}, future work should re-examine what role depth dependent dust, molecular, and continuum opacity may play, especially in setting the terminal velocity of the stellar wind as this is ill-defined under a Parker-like pressure driven wind (see also the discussion in Sect. \ref{sec:mdot_noniso}).
For the remainder of this paper, however, given the highly promising results throughout, we proceed with turbulent pressure supplemented only by the marginal additional contribution of a constant $\Gamma_\mathrm{Ros}$.
%alone.

%\section{Correction factor from numerical steady-state hydrodynamic solutions}

\section{Numerically determined mass-loss rate from steady-state, nonisothermal, dynamical winds}\label{sec:mdot_noniso}

\subsection{Iteration procedure to find a unique mass-loss rate}\label{sec:noniso_iter}

%\begin{itemize}
%    \item full eom
%    \item energy equation given by T-Lucy
%    \item Otherwise, the idea is the same but now with iteration to convergence
%    \item Gives internally consistent density, velocity, temperature, etc.
%    \item Show characteristic structure
%    \item Discussion
%\end{itemize}

While the simplified, isothermal treatment has the benefit of being purely analytic, we can also treat the non-isothermal wind structure in a numerical approach.
To do so, we use the spherically symmetric generalization to the plane parallel gray atmosphere introduced by \cite{Luc71}
%assume the grey opacity temperature structure derived by \cite{Luc71}, namely,

\begin{align}
T^4 &= \Teff^4\left(W + \frac{3}{4} \int_{R_\ast}^r \kappa\, \rho\, \left(\frac{R_\ast}{r}\right)^2 dr\right) \label{eqn:lucy_gray_integral} \\
&= \Teff^4\left(W + \frac{3}{4}\tau\right)\;,
\end{align}
where $W\equiv (1-\sqrt{1-(R_\ast/r)^2}\,)/2$ is the geometric dilution factor.
The second equality again uses the spherically modified optical depth as given by Eq. \ref{eqn:sph_mod_tau}.
For low optical depth material, and for distances away from the stellar surface this temperature structure behaves like $T \propto 1/\sqrt{r}$ as is generally assumed for the dust free regions near red supergiants \citep[e.g.][]{DecHon06}.
Accounting for this temperature structure in Equations \ref{eqn:cons_mass_iso} and \ref{eqn:cons_mom_iso} while still maintaining the assumption that $\vturb$ is constant leads to the system of non-isothermal equations 
%\dylan{somewhat mixed levels of combination here. Either include drho/dr in the second equation or eliminate d tau/dr in favor of kappa mdot/(4 pi v r**2) (Rstar/r)**2. Possibly even fully combine to give one giant equation only for dv/dr?}

\begin{align}
&v\,\ddr{\rho} + \rho\, \ddr{v} + \frac{2\, \rho\, v}{r} = 0 \label{eqn:cons_mass}\\
&v\, \ddr{v} = -\frac{\cs^2 +\vturb^2}{\rho}\,\ddr{\rho}
- \frac{G\, M_\ast\,(1-\Gamma)}{r^2} - \frac{k_\mathrm{B}}{m_\mathrm{H}}\,\ddr{T}  \label{eqn:cons_mom}\\
&\ddr{T^4} = \Teff^4 \left(\ddr{W} + \frac{3}{4} \kappa \rho \left(\frac{R_\ast}{r}\right)^2\right)\label{eqn:lucy_gray_differential}\;.
\end{align}
Note that, by combining this system of differential equations into a single equation to eliminate $\partial \rho/\partial r$ and $\partial T/\partial r$ in Equation \ref{eqn:cons_mom}, we can see that these equations still contain the same critical point where $v^2 = \cs^2 + \vturb^2$, although now with $\cs$ a function of $r$.

In order to solve these differential equations, we begin by assuming initial velocity and density profiles

\begin{align}
    v(r) &= \frac{\sqrt{\cs^2+\vturb^2}}{1 - \Rpmod/r}\left(1-\frac{R_\ast}{r}\right)\label{eqn:initial_v} \\
    \rho(r) &= \frac{\dot{M}}{4\, \pi\, v(r)\, r^2}\label{eqn:initial_rho}\;.
\end{align}
As boundary conditions, we fix $R_\ast$ and impose that $T(R_\ast) = \Teff$.
These are then sufficient conditions to begin the following iteration procedure. %\jon{Actually, I'd say an even more important thing here is the boundary conditions; for the below I take it you assume a *fixed* lower boundary Rstar? Is anything more fixed? Please clarify.}  

\begin{enumerate}
    \item A radial grid is built up inward (toward $R_\ast$) and outward (away from $R_\ast$) from the critical point, $\Rpmod$, as determined in the prior iteration (or the initial conditions for the first iteration).
    Radial points are spaced by $H(r)/3$ to properly resolve the variations in all hydrodynamic quantities.
    \item The prior structures in velocity, density, and temperature are interpolated onto the new radial grid.
    \item Gradients of all necessary quantities for Eqns. \ref{eqn:cons_mass}, \ref{eqn:cons_mom}, and \ref{eqn:lucy_gray_differential} are computed at each grid point.
    At the critical point, $r = \Rpmod$, where the velocity gradient is ill-defined, l'Hopital's rule is used.
    \item Using these gradients, Eqns. \ref{eqn:cons_mass}, \ref{eqn:cons_mom}, and \ref{eqn:lucy_gray_differential} are numerically solved at each point on the radial grid inward and outward from the critical point, $\Rpmod$.
    \item The resulting velocity, density, and temperature as functions of radius are used to begin the process again from Step 1.
\end{enumerate}
This loop is repeated until velocity, density, and temperature at each point agree from one iteration to the next to better than 1\%.
At this point, the total optical depth of the wind is computed at the current mass loss rate.
As the stellar radius is defined to be at $\tau(R_\ast) = 2/3$, the stellar mass-loss rate is re-computed according to 

\begin{equation}
    \dot{M}_\mathrm{new} = \dot{M}_\mathrm{old} \frac{2/3}{\tau_\mathrm{old}(R_\ast)}\;.
\end{equation}
As optical depth is proportional to density, which is in turn proportional to $\dot{M}$, a constant change in mass-loss rate and thus density at every point has the effect of forcing $\tau(R_\ast) = 2/3$, assuming $v(r)$ is unchanged.
As changing $\rho(r)$ will certainly change $v(r)$, however, the final velocity profile is combined with the updated mass loss as new initial conditions to restart the loop described above.
%the new mass loss rate is used to compute a new initial condition from Eqns. \ref{eqn:initial_v} and \ref{eqn:initial_rho} to restart the loop described above. 
%\jon{Actually, I think you might make this significantly more efficient by starting from the v structure in the previous iteration-cycle instead. Is this something you've tried?} 
%\dylan{I have tried this. The models run fast enough in either case that optimization was not a big concern. However, I have to re-run the grid again (stupid mistake I didn't catch, see below) so I have made this change since you are correct that it makes a very slight speed up}
This outer iteration to update mass-loss rate is repeated until the update in $\dot{M}$ is also less than 1\%, at which the final mass loss rate is returned as $\dot{M}_\mathrm{num}$.
Note that for the remainder of the paper we include the subscript ``num'' to differentiate this numerically determined, non-isothermal mass-loss rate from the isothermal, analytic one derived in Sect. \ref{sec:mdot_iso}, which we henceforth call $\dot{M}_\mathrm{an}$.

\begin{figure*}
\centering
\begin{subfigure}{0.32\textwidth}
\includegraphics[width=\textwidth]{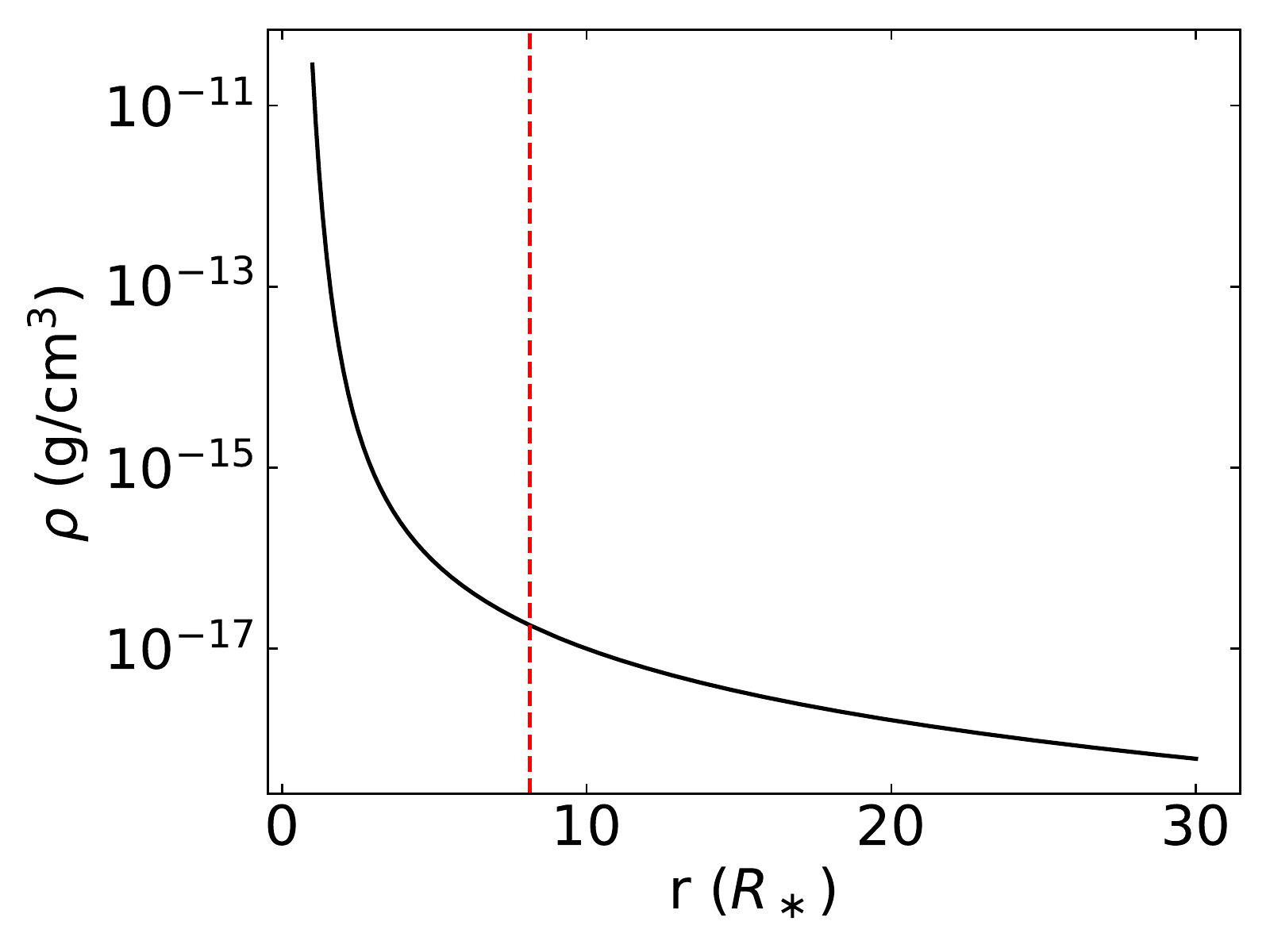}
\end{subfigure}
\begin{subfigure}{0.32\textwidth}
\includegraphics[width=\textwidth]{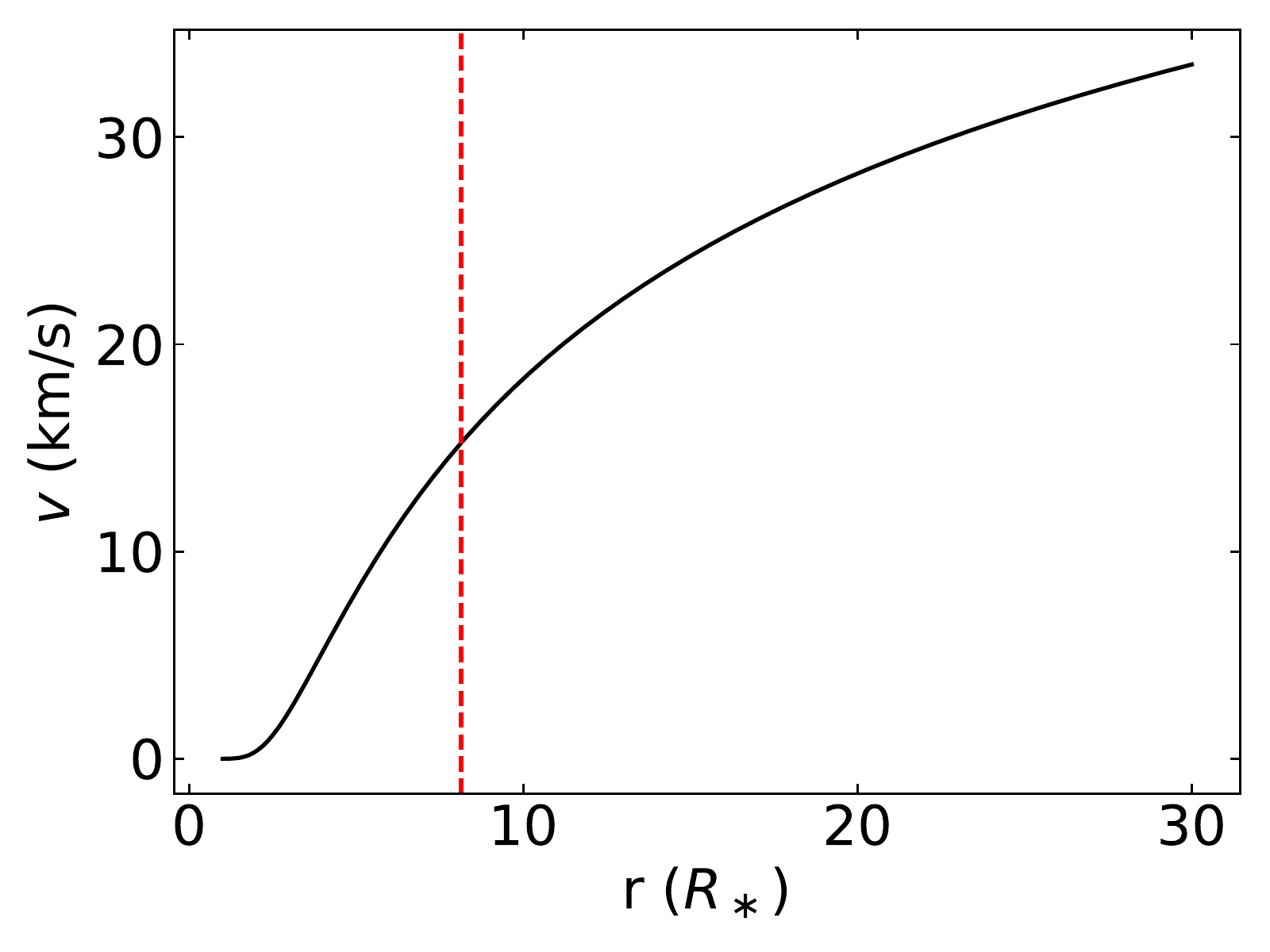}
\end{subfigure}
\begin{subfigure}{0.32\textwidth}
\includegraphics[width=\textwidth]{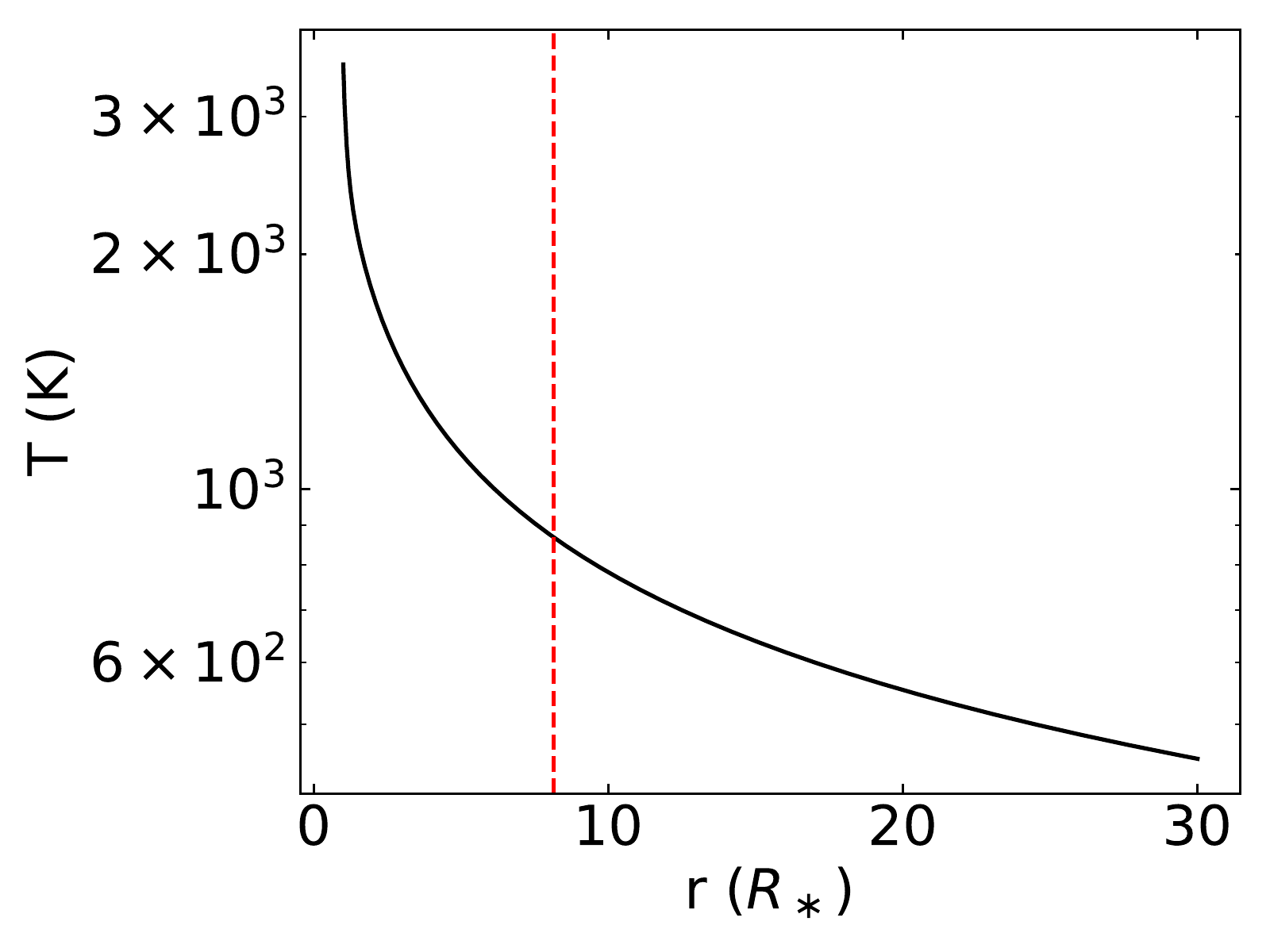}
\end{subfigure}
\caption{\label{fig:sample_structure} Radial dependence of density (left), velocity (center), and temperature (right) for the non-isothermal case with $M_\ast = 10 \Msun$, $R_\ast = 500 \Rsun$, $\Teff = 3500$ K, $\vturb = 15$ km s$^{-1}$, and $\kappa = 0.01$ cm$^2$ g$^{-1}$. The red dashed line denotes the modified Parker radius.}
\end{figure*}

\begin{figure}
\centering
\includegraphics[width=0.48\textwidth]{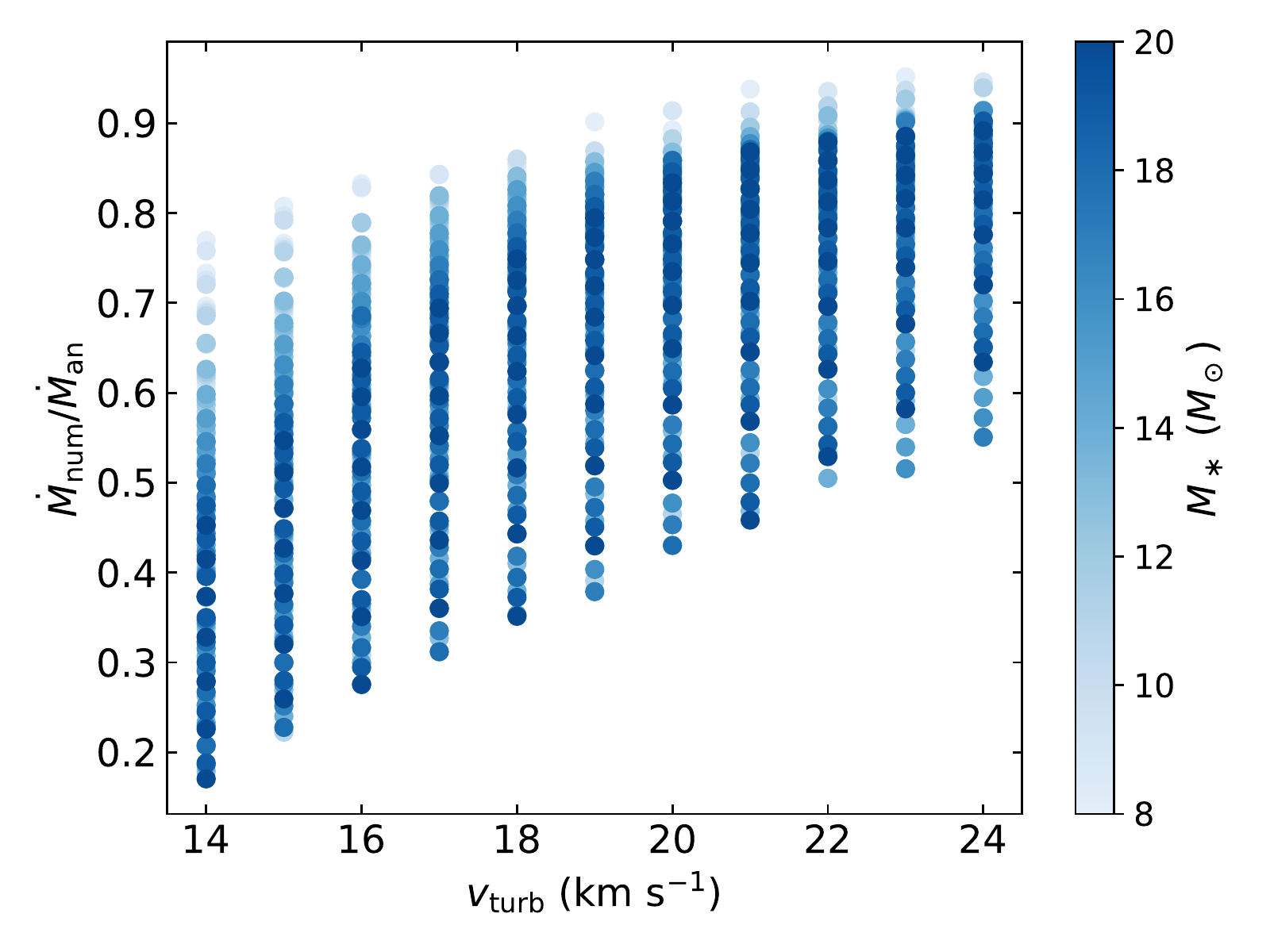}
\caption{\label{fig:corr_factor}
Ratio of $\dot{M}$ calculated numerically for a non-isothermal situation to the analytic $\dot{M}$ calculated for an isothermal wind plotted as a function of $\vturb$ and colored by $M_\ast$, varying $R_\ast$, $M_\ast$, and $v_{\rm turn}$. The effective temperature and Rosseland opacity are held fixed at $\Teff = 3500$ K and $\kappa_\mathrm{R} = 0.01$ cm$^{2}$ g$^{-1}$.} 
\end{figure}

Here it is again interesting to examine our ``typical'' model with $M_\ast = 10 \Msun$, $R_\ast = 500 \Rsun$, $\Teff = 3500$ K, $\vturb = 15$ km s$^{-1}$, and $\kappa = 0.01$ cm$^2$ g$^{-1}$.
Using the method described above, we find that it has a non-isothermal mass-loss rate of $\dot{M}_\mathrm{num} = 4.4 \times 10^{-7} M_\odot$ yr$^{-1}$, approximately half the analytic rate.
We can also take this opportunity to plot the radial profiles of density, velocity, and temperature as shown in Figure \ref{fig:sample_structure}.
In all panels we use a vertical, red, dashed line to denote the numerically determined $\Rpmod$.
As was the case in Sect. \ref{sec:mdot_derivation}, optical depth is predominantly accumulated below $\Rpmod$ such that the mass-loss rate in each simulation is 
%uniquely set 
primarily determined by integration over only these near star regions.
However, like the isothermal, thermal pressure-driven Parker wind, the total driving energy available to the wind diverges as the outer radius goes to infinity because $\vturb$ is constant, such that $v_\infty$ is not uniquely determined by such a model.
While the terminal velocity of the wind will depend on the details of the radial $\vturb$ profile, radial temperature profile, and any additional forces acting on the outer wind (for instance dust opacity), it is promising that these simulations already generate velocities at 30 $R_\ast$ comparable to typically inferred values of $v_\infty$ for red supergiants.

While we here present the mass-loss rate from a simplified gas temperature structure arising from radiative equilibrium, in reality red supergiant winds may have regions that are heated by shocks and dissipation of turbulence \citep[see, e.g.][]{SchZwa00}.
In general, such shock heating and dissipation is not inferred to generate wind temperatures throughout the full wind acceleration region significantly in excess of the stellar effective temperature, nor should the wind be expected in general to cool below radiative equilibrium.
Therefore, the simplified cases we present here of an isothermal wind at the stellar effective temperature and a wind with temperature set according to the spherically modified gray atmosphere can be roughly taken as thermal structures leading to, respectively, upper and lower limits to the mass-loss rate.
For individual objects where more is known about the temperature profile, however, the method presented in this section can be directly applied by replacing the temperature information in Equations \ref{eqn:lucy_gray_integral} and \ref{eqn:lucy_gray_differential}.
As a sample case, let us take the chromospheric temperature profile inferred by \cite{OGoHar20} for Antares.
In these observations, gas temperature is inferred to decrease from the stellar effective temperature to a minimum around $1.35 R_\ast$ before coming back up to a maximum approximately equal to the stellar effective temperature around $2.5 R_\ast$, and then finally falling off outward from there \citep[see][their Figure 3]{OGoHar20}.
Inserting this temperature profile into our model with all other parameters still as defined in our ``typical'' model above returns a mass-loss rate $7.7\times10^{-7} M_\odot$ yr$^{-1}$. As anticipated, this falls between the limiting mass-loss rates from an isothermal model at the stellar effective temperature ($8.5\times10^{-7} M_\odot$ yr$^{-1}$) and a model using the \cite{Luc71} spherically modified gray atmosphere ($4.4\times10^{-7} M_\odot$ yr$^{-1}$).

\subsection{Non-isothermal correction factor}\label{sec:noniso_correction}

To investigate the role of the radial decline in temperature on the mass-loss rate, we run a grid of steady-state models varying stellar mass, stellar radius, and turbulent velocity and then plot the ratio $\dot{M}_\mathrm{num}/\dot{M}_\mathrm{an}$.
For this test we fix $\Teff = 3500$ K, motivated by the only modest dependence of $\dot{M}$ on this property (see Figure \ref{fig:mdot_with_stellar}), and continue to use $\kappa_R = 0.01$ cm$^{2}$ g$^{-1}$.
%\dylan{variations on other parameters to expand the grid? kappa and Teff?}
As shown by Figure \ref{fig:corr_factor}, this correction factor is itself a function of the model parameters that we can fit.
We choose two methods of doing so, one where we fit the product of a constant normalization and power-laws over stellar mass, stellar radius, and turbulent velocity individually (hereafter Method 1), and one where we apply our knowledge that the dominant competition in getting material off the star is between gravity and turbulent pressure to replace the power-law fits over each parameter independently with a single power-law fit over $\vturb/v_\mathrm{esc}(M_\ast,R_\ast)$ (hereafter Method 2).
%We choose two methods of doing this, one where we fit over each parameter individually yielding

%\begin{equation}\label{eqn:corr_fit_1}
%\left(\frac{\dot{M}_\mathrm{num}}{\dot{M}_\mathrm{an}}\right)_1 = 10^{-0.24} %\left(\frac{M_\ast}{12\;M_\odot}\right)^{- 0.68}\left(\frac{R_\ast}{1000\;R_\odot}\right)^{0.67}\left(\frac{\vturb}{15\;\mathrm{km}\;\mathrm{s}^{-1}}\right)^{2.13}\;,
%\end{equation}
%and a second where we apply our knowledge that the dominant competition in getting material off the star will be between gravity and turbulent pressure to fit over $\vturb/v_\mathrm{esc}(M_\ast,R_\ast)$ such that

%\begin{equation}\label{eqn:corr_fit_2}
%\left(\frac{\dot{M}_\mathrm{num}}{\dot{M}_\mathrm{an}}\right)_2  =10^{0.00} %\left(\frac{\vturb/(15\; %\mathrm{km}\;\mathrm{s}^{-1})}{v_\mathrm{esc}(M_\ast,R_\ast)/(50\;\mathrm{km}\;\mathrm{s}^{-%1})}\right)^{1.85}\;.
%\end{equation}
%The subscripts in Equations \ref{eqn:corr_fit_1} and \ref{eqn:corr_fit_2} will be used to refer to these as Methods 1 and 2 respectively.

%\begin{table}
%\centering
%\caption{\label{tab:fit_compare} Comparison of the fit methods}
%\begin{tabular}{l|c|c}
%& Method 1 & Method 2 \\
%\hline
%$\chi^2$ & 3.54 & 6.45 \\
%$\langle$ Error $\rangle$ & 5.6$\times 10^{-2}$  & 8.1$\times 10^{-2}$ \\
%$\sigma_\mathrm{Error}$ & 5.4$\times 10^{-2}$ & 7.5 $\times 10^{-2}$ \\
%max(Error) & 0.49 & 0.57 \\
%\end{tabular}
%\end{table}

\begin{table}
\centering
\caption{\label{tab:fit_compare} Comparison of the fit methods}
\begin{tabular}{l|c|c}
\hline
\hline
& Method 1 & Method 2 \Tstrut \Bstrut \\
\hline
$\chi^2$ & 6.73 & 13.11 \Tstrut \\
$\langle$ Error $\rangle$ & 0.091  & 0.118 \\
$\sigma_\mathrm{Error}$ & 0.077 & 0.119 \\
max(Error) & 0.81 & 1.01 \\
\end{tabular}
\end{table}

As figures of merit to compare the two fits, we use $\chi^2$, the mean and standard deviation of the difference between the fit formula and the actual ratio $\dot{M}_\mathrm{num}/\dot{M}_\mathrm{an}$, and the maximum error between the fit and $\dot{M}_\mathrm{num}/\dot{M}_\mathrm{an}$.
The comparison of the best models from each method is summarized in Table \ref{tab:fit_compare}.
Given that method 1 uses four free parameters (a mean and three power-law indices) versus the two free parameters used by method 2 (a mean and one power-law index), and that the figures of merit we have selected to compare between the two methods are comparable, we propose that the two parameter fit from method 2 should be preferable moving forward.
Therefore, the non-isothermal correction factor is
\begin{equation}\label{eqn:corr_fit_2}
\left(\frac{\dot{M}_\mathrm{num}}{\dot{M}_\mathrm{an}}\right)  = \left(\frac{\vturb/(17\; \mathrm{km}\;\mathrm{s}^{-1})}{v_\mathrm{esc}(M_\ast,R_\ast)/(60\;\mathrm{km}\;\mathrm{s}^{-1})}\right)^{1.30}\;,
\end{equation}
where we have omitted the leading constant as the best fit for this constant was 10$^{0.00} = 1.0$.

%\jon{There seems to be something funny about this formula. I don't understand what the "2" on lhs is, neither what the meaning of $10^(0.00)$ on rhs is? Please check.}

%\dylan{The ``2'' was a hold over from a prior draft. The $10^{0.00}$ is the fit to the leading constant term which is log10(const) = 0.00 to within 0.01. Since I still do fit two parameters here, I am not sure how else to denote this...}

Note from Table \ref{tab:fit_compare} that this fit is quite good, with the correction factor recovered to about a factor of two in the worst case for the two parameter fit.
Therefore we present the combination of Equation \ref{eqn:corr_fit_2} with Equations \ref{eqn:Rpmod}, \ref{eqn:mdot_an}, and \ref{eqn:rhoRp} as a predictive, theoretical mass-loss rate scaling for red supergiants.

\section{Comparison of the theoretical mass-loss rate scaling with empirical rates and observed mass-loss}\label{sec:comp_w_emp_and_obs}

Now that we have this theoretical, predictive mass-loss rate as a function of stellar parameters, a natural next step is to compare this with empirical fits to red supergiant mass loss and individual observations.
%the current state of the art in red supergiant mass-loss predictions, as well as with observed mass-loss rates. \jon{Don't understand what you mean here above. I thought there weren't really any theoretical predictions for RSGs? Please clarify (I know just below it seems you do so, but it reads funny with your statement above).} 
%Given that previous theoretical predictions have been unsatisfactory, empirical fits constitute the state of the art in theory of red supergiant mass loss predictions.
Given that no systematic theoretical predictions for RSG mass loss exist at the present, such empirical fits constitute the state of the art regarding RSG mass loss in applications such as stellar evolution.
We review the empirical rates implemented in the stellar evolution code MESA, as well as the recently published rates from \cite{BeaDav20} in Sect. \ref{sec:emp_rates} in order to build up a representative sample.
In Sect. \ref{sec:comp_w_emp_rates} we then compare these empirical rates to our theoretical ones.
Here we highlight that the empirical rates in general do not depend on all three fundamental stellar parameters (i.e. $M_\ast$, $R_\ast$, and $L_\ast$) while the theoretical mass-loss rates presented here does.
Therefore Sect. \ref{sec:comp_w_emp_rates} also includes a discussion of the scalings of stellar luminosity, mass, and radius with one another that we use to perform the comparison of our rates with the empirical ones.
%This includes a discussion of the scalings of stellar luminosity, mass, and radius with one another as these are necessary to compare the theoretical rates predicted from all three of these to empirical rates which in some cases are fits to only two of these parameters.
%\jon{LONG complicated sentence above. Could you perhaps break off by, at least, some coma or so. Or clarify and make two sentences.} 
Finally, in Sect. \ref{sec:comp_w_obs} we compare our predicted rates with observationally inferred mass-loss rates for a selection of red supergiants.

\subsection{Background on empirical rates used so far}\label{sec:emp_rates}

The current version of the stellar evolution code \code{MESA} \citep{Paxton11, Paxton13, Paxton15, Paxton18, Paxton19} provides built-in access to three empirical mass-loss rates for red supergiants.
These come from \cite{deJNie88}

%\begin{equation}
%    \log_{10}\left(\dot{M}\right) = - 8.158 + 1.769 \log_{10}\left(\frac{L_\ast}{\Lsun}\right) - 1.676 \log_{10}\left(\Teff\right)\;,
%\end{equation}

\begin{equation}
    \dot{M} = 10^{- 8.158} \left(\frac{L_\ast}{\Lsun}\right)^{1.769} \left(\Teff\right)^{- 1.676}\;,
\end{equation}
\cite{NiedeJ90}

%\begin{equation}
%    \log_{10}\left(\dot{M}\right) = -14.02 + 1.24 \log_{10}\left(\frac{L_\ast}{\Lsun}\right) + 0.16 \log_{10}\left(\frac{M_\ast}{\Msun}\right) + 0.81 \log_{10}\left(\frac{R_\ast}{\Rsun}\right)
%\end{equation}
\begin{equation}
    \dot{M} = 10^{-14.02} \left(\frac{L_\ast}{\Lsun}\right)^{1.24} \left(\frac{M_\ast}{\Msun}\right)^{0.16} \left(\frac{R_\ast}{\Rsun}\right)^{0.81}
\end{equation}
and \cite{LooCio05}

%\begin{equation}
%    \log_{10}\left(\dot{M}\right) = -5.65 + 1.05\log_{10}\left(\frac{L_\ast}{10^4 \Lsun}\right) - 6.3\log_{10}\left(\frac{\Teff}{3500}\right)\;.
%\end{equation}
\begin{equation}
    \dot{M} = 10^{-5.65} \left(\frac{L_\ast}{10^4 \Lsun}\right)^{1.05} \left(\frac{\Teff}{3500}\right)^{- 6.3}\;.
\end{equation}
To these we add the recently published empirical rate from \cite{BeaDav20}

\begin{equation}
    \dot{M} = 10^{-26.4 - 0.23 M_\ast/\Msun} \left(\frac{L_\ast}{L_\odot}\right)^{4.8}\;,
\end{equation}
as the authors identify that this rate is notably lower than those previously published, a fact that we use to provide an indication to the range of possible observationally inferred rates.
While these four are far from the only empirical mass-loss prescriptions that are used, they provide a representative sample.
For a review of a variety of other empirical rates, see, for example, \cite{MauJos11}.

\subsection{Comparison of the theory model to empirical rates}\label{sec:comp_w_emp_rates}

\begin{figure}
    \centering
    \includegraphics[width=0.48\textwidth]{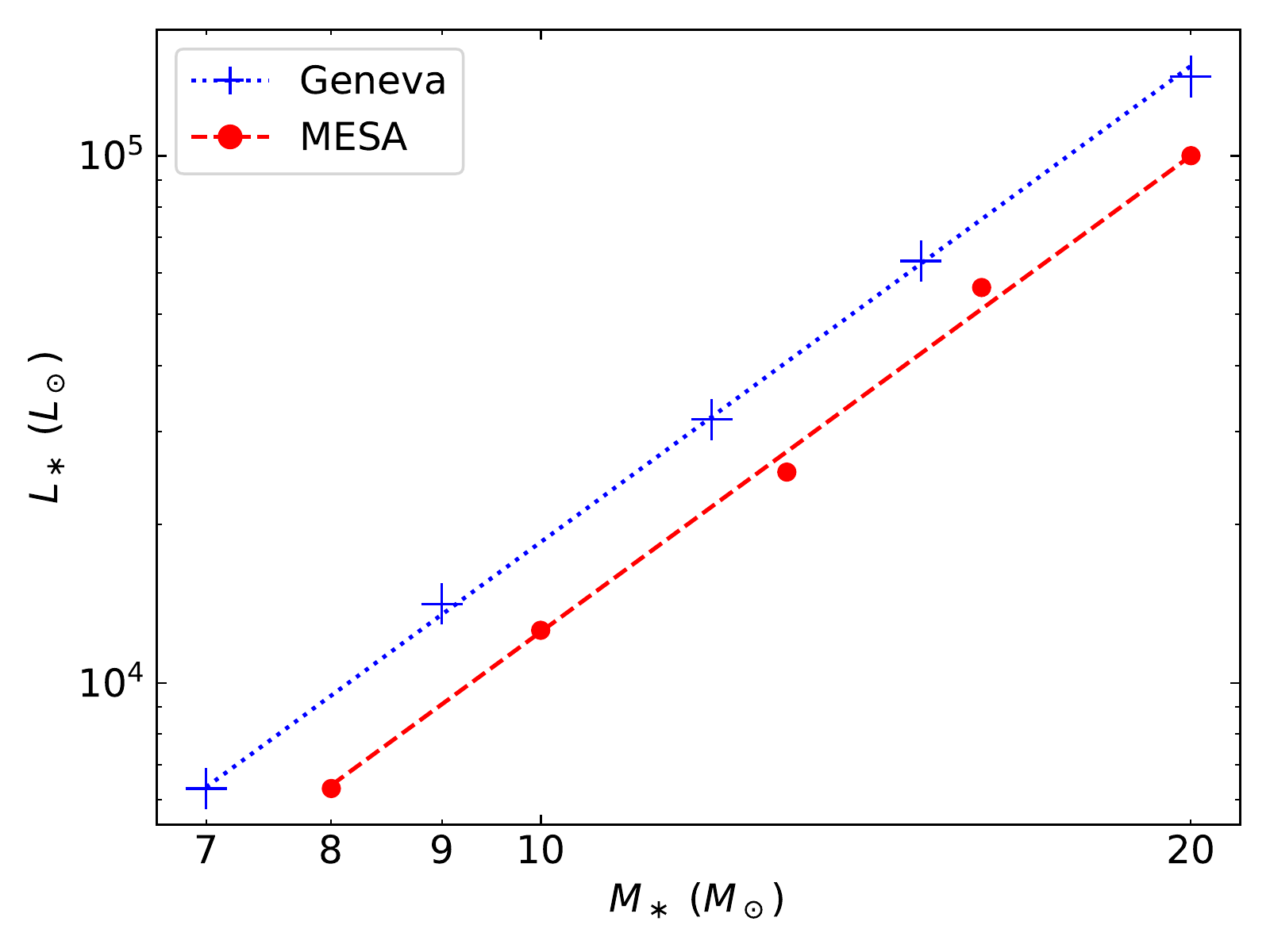}
    \caption{Luminosity as a function of mass from red supergiants in \cite{Paxton11} and \cite{EksGeo12}. The overplotted curves assume $L_\ast = f M_\ast^3$ with $f = 12.5$ for the MESA data and $f=18.5$ for the Geneva data.}
    \label{fig:mass_lum_rel}
\end{figure}

As mentioned in the prior subsection, it is common for empirical rates to be provided as functions that do not include a dependency on all three fundamental stellar parameters (i.e. $M_\ast$, $R_\ast$, and $L_\ast$, or equivalently $M_\ast$, $L_\ast$, and $\Teff$).
For the theoretical rates, however, all three parameters are needed. 
%such that a one-to-one comparison is not possible without mass-luminosity-radius relations for red supergiants.
To obtain these, we consider stellar evolution tracks for nonrotating stars from both \code{MESA} \citep{Paxton11} and \code{GENEC} \citep{EksGeo12}.
Note that \code{GENEC} uses a somewhat different prescription for the mass-loss rate than what is discussed above for \code{MESA}.
For $M_\ast \leq 12 M_\odot$ \code{GENEC} adopts the \cite{Rei75,Rei77} rates, whereas for $M_\ast \geq 15M_\odot$ \code{GENEC} uses \cite{deJNie88} for $\log_{10}(\Teff) > 3.7$ and a linear fit to data from \cite{SylSki98} and \cite{LooGro99} otherwise.
Despite these differences in mass-loss prescription, comparison of \cite{Paxton11} and \cite{EksGeo12} shows that in both cases the red supergiant branch roughly follows $L_\ast = f M_\ast^3$ as shown by Figure \ref{fig:mass_lum_rel} with $f=12.5$ for MESA and $f=18.5$ for Geneva.
Therefore, as a compromise we chose $f=15.5$ for our mass-luminosity relation to compare with the empirical rates.
As a simplifying assumption, we continue to hold $\Teff$ fixed at 3500 K as was done in Sect.~\ref{sec:mdot_noniso}, so that
%This means that that 
$L_\ast \propto R_\ast^2$. 
%through $L_\ast = 4 \pi R_\ast^2 \sigma \Teff^4$.

\begin{figure}
\centering
\includegraphics[width=0.48\textwidth]{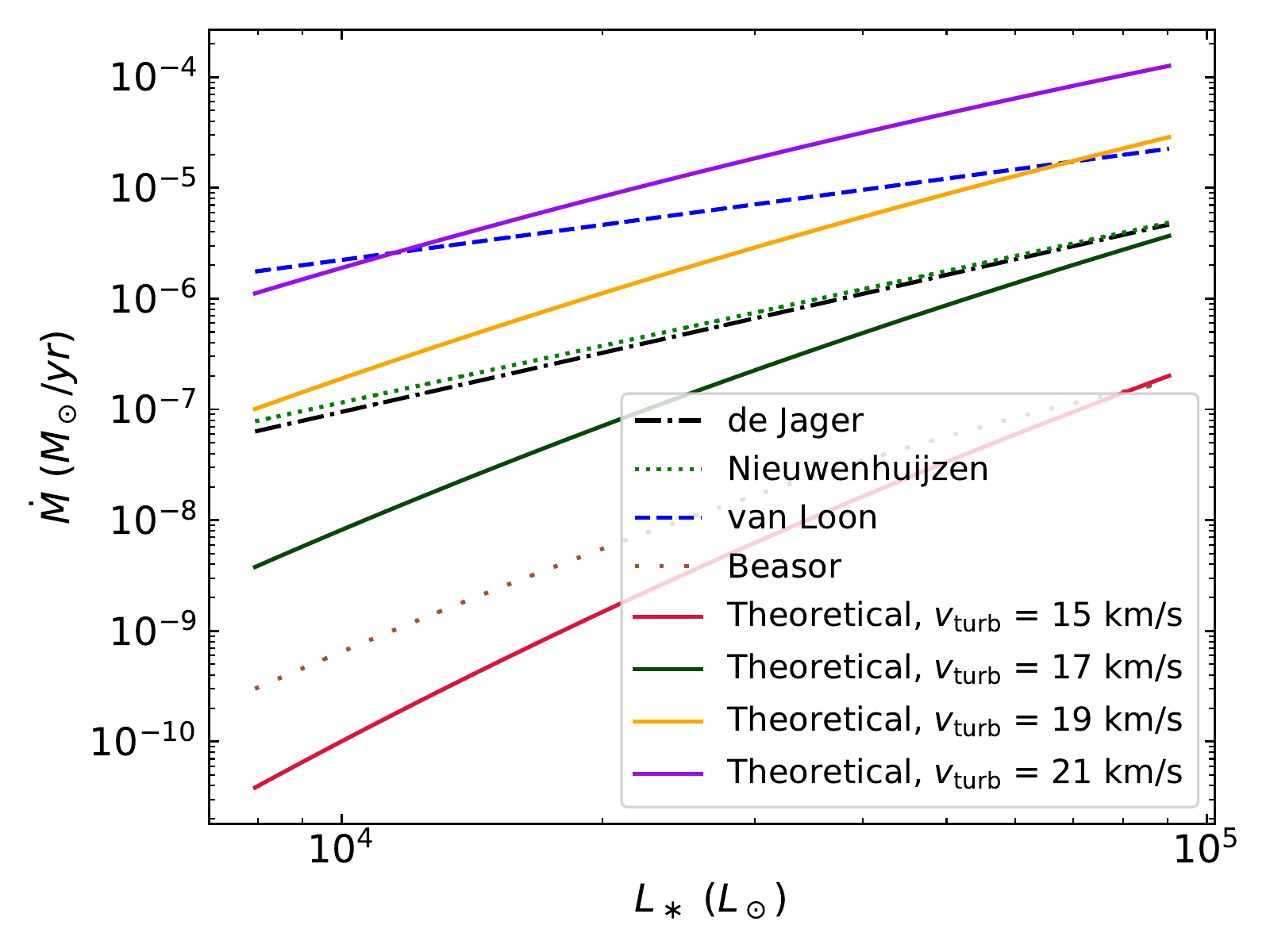}
\caption{\label{fig:theory_vs_empirical}
An overplot of three of the theoretical mass-loss rate scalings used in MESA with the predicted mass-loss rate from the theoretical scaling. Four values of $\vturb$ are used as noted in the plot label.
}
\end{figure}

Using these mass-radius-luminosity relationships, we can now compare the empirical and theoretical rates as shown by Figure \ref{fig:theory_vs_empirical}.
As the empirical rates differ from one another by several orders of magnitude at all luminosities, we select four values of $\vturb$ from 15 to 21 km s$^{-1}$ for this comparison.
Despite this scatter between individual empirical rates and in turn between the theoretical rates using different turbulent velocities, the theoretical rates are broadly consistent with the empirical ones for reasonable values of $\vturb$ \citep[see, e.g.][]{JosPle07,OhnWei17}.
Further examining this comparison, one notes that the empirical rates have shallower slopes than the theoretical rates.
This may suggest that there is a trend of decreasing $\vturb$ for more massive, more luminous, and larger radius red supergiants.
Alternatively, as is suggested by the mismatch in both value and slope of the empirical rates with one another, this shallower slope may simply reflect some missing physics when inferring and calibrating these empirical rates or potential biases in the observed sample.

We note that the mass-luminosity-radius relations that we use here may not be appropriate to Galactic stars starting out their lives rapidly rotating, as these tend to evolve to higher luminosities for the same initial mass \citep[see, e.g.][]{HegLan00,MeyMae00,BrodeM11}.
Given, however, the very similar effective temperatures of these stars on the red supergiant branch this is consistent with these stars evolving to have larger radii, lower surface gravities, and higher mass-loss rates if all other physics remains the same.
This further suggests that two stars with different masses can reach the same point on the HR diagram with different mass-loss rates which would confound any mass-loss rate predictions omitting a dependence on $M_\ast$. 
In fact the general treatment that mass-loss rates are enhanced by rotation would go in the direction of only further exacerbating this discrepancy \citep[see, e.g.][]{FriAbb86,Lan97,HegLan98,MaeMey00}.
However, as further discussed in Sect. \ref{sec:outlook}, these predictions of increased mass-loss rate with increased rotation rate are based on fits to the theoretical predictions of \cite{FriAbb86}.
This work addresses only the wind mass-loss rate of hot stars with radiation driven outflows, and has been questioned by subsequent work \citep[e.g.][]{OwoCra96,PetPul00} such that the question of the role of rotation in the mass-loss of red supergiants remains open.

%as is suggested by the empirical rates to reproduce one another, the lower slopes in the these empirical rates may simply be an artifact of attempting to fit many objects with different parameters 
%\jon{OR, and more likely at least in my opinion..., that these empirical scaling relations are quite off..} 

%\dylan{For comparison, can assume horizontal evolution on HRD to know mass-luminosity relation. Then assuming Teff 3500 K gives radius-luminosity relation.}

\subsection{How does the theory model compare to what is observationally inferred?}\label{sec:comp_w_obs}

\begin{table*}
    \centering
    \begin{threeparttable}
        \caption{Table of observed parameters from \cite{JosPle07} with the addition of Antares as observed by \cite{OhnHof13,OhnWei17}, including $\vturb$ as required by theory to recover the mass-loss rate keeping all other parameters fixed. \label{tab:obs_theory_comparison}}
        \begin{tabular}{cccccccc}
            \hline
            \hline
            Number & Name & Mass & $\Teff$ & Radius & $\dot{M}_\mathrm{gas}$\tnote{a} & ${\vturb}_\mathrm{,Obs}$ & ${\vturb}_\mathrm{,Theory}$ \Tstrut \\
           & & $\Msun$ & K & $\Rsun$ & $10^{-7}\;M_\odot$ yr$^{-1}$ & \multicolumn{2}{c}{km s$^{-1}$} \Bstrut \\
            \hline
            1 & $\alpha$ Ori & 15 & 3780 & 589 & 5.0 & 19 & 17\Tstrut \\
            2 & V466 Cas & 12 & 3780 & 331 & 0.5 & 12 & 19 \\
            3 & AD Per & 12 & 3720 & 457 & 2.0 & 21 & 17 \\
            4 & FZ Per & 12 & 3920 & 324 & 1.75 & 16 & 20 \\
            5 & BD+243902 & 15 & 4240 & 427 & 7.25 & 23 & 21 \\
            6 & BI Cyg & 20 & 3720 & 851 & 10.25 & 23 & 16 \\
            7 & BC Cyg & 20 & 3570 & 1230 & 8.0 & 22 & 13 \\
            8 & RW Cyg & 20 & 3920 & 676 & 8.25 & 20 & 19 \\
            9 & SW Cep & 9 & 3570 & 234 & 11.5 & 24 & 23 \\
            10 & $\mu$ Cep & 25 & 3750 & 1259 & 3.75 & 23 & 14 \\
            11 & ST Cep & 9 & 4200 & 174 & 6.25 & 23 & 26 \\
            12 & TZ Cas & 15 & 3670 & 646 & 9.5 & 17 & 17 \\
            13 & Antares & 12.7\tnote{b} & 3660\tnote{b} & 680\tnote{b} & 20.0\tnote{c} & 20\tnote{d} & 15\Bstrut\\
            \hline
        \end{tabular}
        \begin{tablenotes}
            \item[a] A constant gas to dust mass ratio 250 has been assumed to convert $\dot{M}_\mathrm{dust}$ as compiled by \cite{JosPle07} to $\dot{M}_\mathrm{gas}$.
            \item[b] \cite{OhnHof13}
            \item[c] \cite{BraBaa12}
            \item[d] \cite{OhnWei17}
        \end{tablenotes}
    \end{threeparttable}
\end{table*}

While the preceding comparison of the empirical rates with theoretical ones helps to generally motivate the quality of the theoretical mass-loss rate predictions, we can go one step further and compare theory and observation for a sample of observed stars.
For this we select the observations compiled by \cite{JosPle07} with the addition of Antares using data from \cite{OhnWei17}.
%, as these authors additionally compiled velocity dispersion measurements which they associate with turbulent motions.
%\dylan{
Note that this limited sample is based on these authors compiling velocity dispersion measurements that we can compare with the $\vturb$ required by our model.
%}
Table \ref{tab:obs_theory_comparison} recapitulates the stellar masses, effective temperatures, and radii from \cite{JosPle07} and \cite{OhnWei17}.
%\dylan{
As the model we present predicts gas mass-loss rates while \cite{JosPle07} report dust mass-loss rates, we apply a constant gas to dust mass ratio $\dot{M}_\mathrm{gas} = 250 \dot{M}_\mathrm{dust}$.
This ratio is itself not particularly well constrained by the existing literature as shown by a sample comparison of the gas mass-loss rates of \cite{deBDec10} with the dust mass-loss rates of \cite{HarMal01} and \cite{Vervan09} for VY CMa, $\mu$ Cep, and $\alpha$ Ori, which returns ratios between $\sim$70 and $\sim$700.
%}
%as directly inferring gas mass-loss rate as is computed by the theoretical model, 
%we take the dust mass loss rates reported by \cite{JosPle07} and apply a constant dust mass fraction 
%of 1\% such that $\dot{M}_\mathrm{gas} = 100 \dot{M}_\mathrm{dust}$.
Finally, we report the velocity dispersion measurement that \cite{JosPle07} and \cite{OhnWei17} associate with turbulent velocity as $\vturbObs$, and the turbulent velocity that would be required for the theoretical model to reproduce $\dot{M}_\mathrm{gas}$ as $\vturbTh$.
For the data from \cite{JosPle07}, we round their reported values to the nearest km s$^{-1}$ as the errors on the original data may be as large as $\pm 2$ km s$^{-1}$.
We report our required $\vturb$ with the same accuracy.

\begin{figure*}
\centering
\begin{subfigure}{0.48\textwidth}
\includegraphics[width=\textwidth]{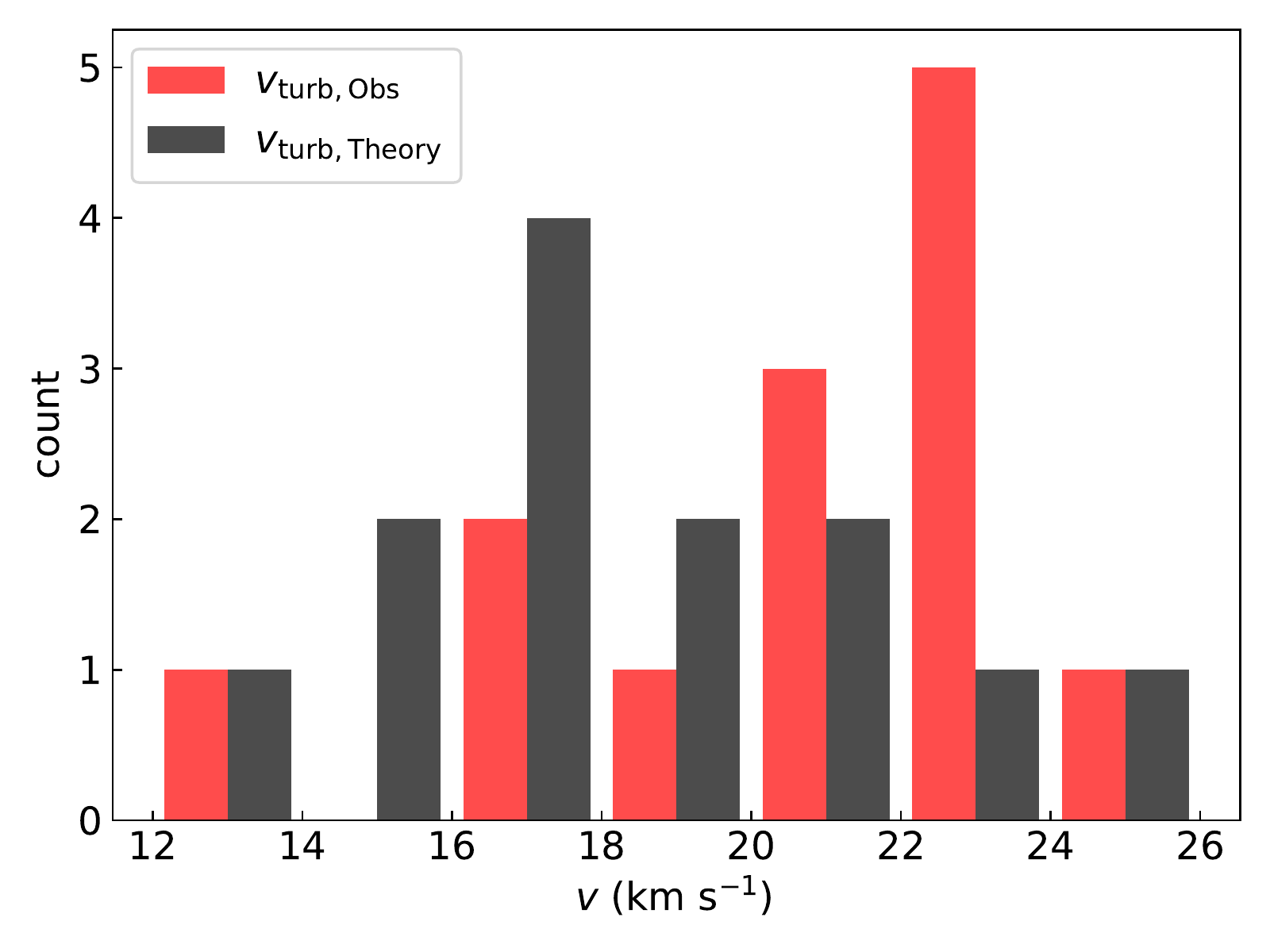}
\end{subfigure}
\begin{subfigure}{0.48\textwidth}
\includegraphics[width=\textwidth]{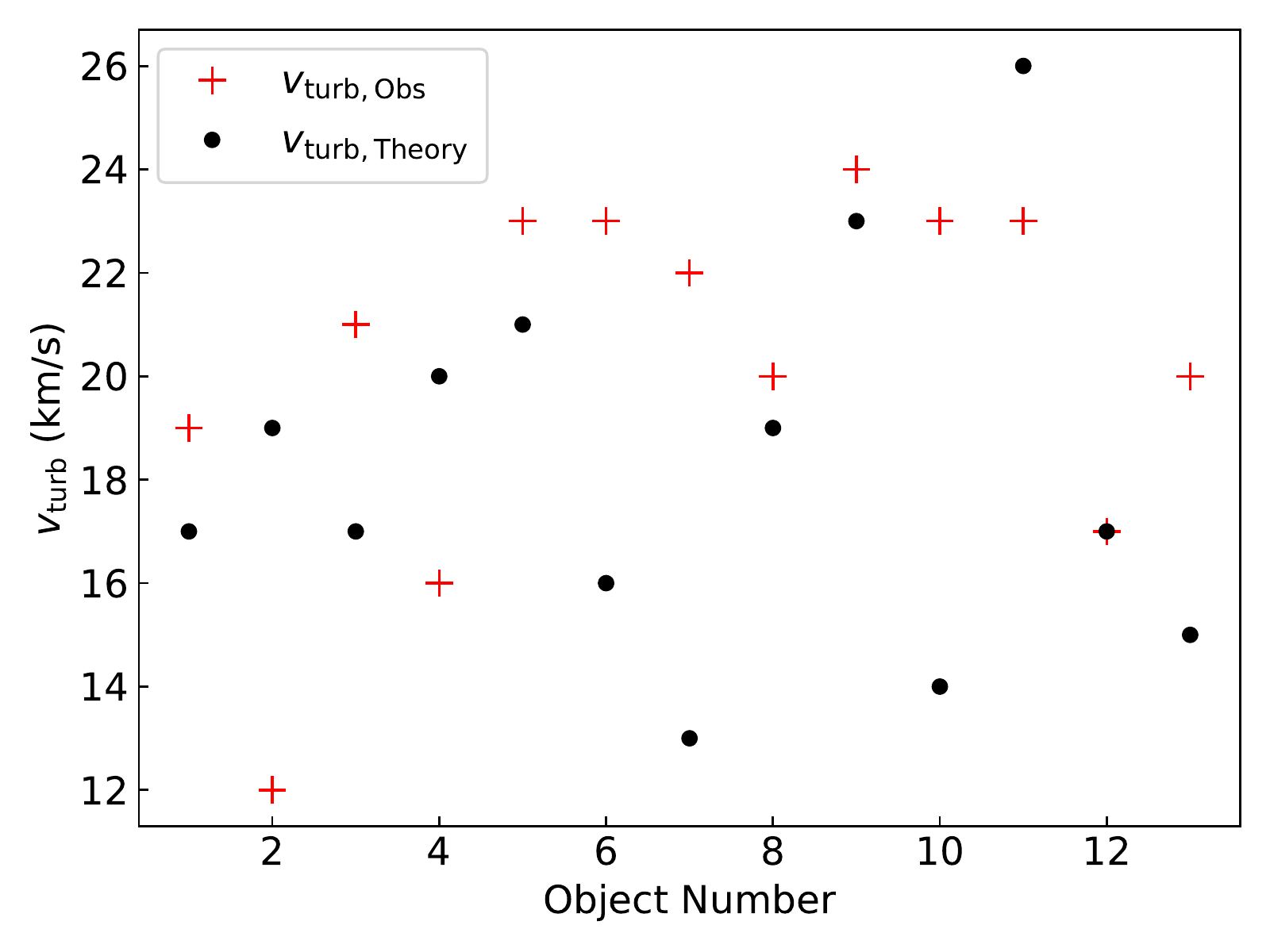}
\end{subfigure}
\caption{\label{fig:vturb_Obs_theory}
Comparisons of $\vturb$ from observations ($\vturbObs$) and from matching corresponding observationally inferred mass-loss rates to the theoretical predictions presented here (see text). The left panel presents a histogram of the distribution while the right panel plots $\vturbObs$ and $\vturbTh$ for each object numbered according to Table \ref{tab:obs_theory_comparison}.}
\end{figure*}

Comparing the mean value of the observed turbulent velocity $\overline{v}_{\rm turb,\,Obs}$ to the mean theoretical turbulent velocity $\overline{v}_{\rm turb,\,Theory}$, the two are consistent with each other over the population to within one standard deviation, as $\overline{v}_\mathrm{turb,\,Obs} = 20.2 \pm 3.4$ km s$^{-1}$ and $\overline{v}_\mathrm{turb,\,Theory}~=~18.2~\pm~3.5$~km~s$^{-1}$.
%\dylan{
Repeating this exercise with the extremal gas to dust mass ratios above also retrieves $\overline{v}_\mathrm{turb}$ values consistent with the $\overline{v}_\mathrm{turb,\,Obs}$ and $\overline{v}_\mathrm{turb,\,Theory}$ values reported here to within one standard deviation, such that this choice has not significantly impacted our results.
%}
These ranges are also generally consistent with the turbulent velocities required to recover the empirical rates in the prior subsection.
Moreover, the overall ranges are consistent with $\vturbObs \in$~[12,~24]~km~s$^{-1}$ and $\vturbTh \in$~[13,~26]~km~s$^{-1}$.
Finally, looking at the distributions of $\vturb$ for both theory and observation, plotted in the left panel of Figure \ref{fig:vturb_Obs_theory}, as well as comparing $\vturbObs$ to $\vturbTh$ for individual objects as plotted in the right panel of Figure \ref{fig:vturb_Obs_theory}, one can see that the majority of turbulent velocities required by theory (10 out of 13) are lower than what is observationally inferred.
This argues that turbulent pressure is an ample driving source for the wind of red supergiants, strongly reinforcing the quality of this model in explaining 
%\jon{
the general behavior of these objects' mass loss.
%} 
%these object's mass loss.

As turbulent velocity is not observationally inferred for many red supergiants, $\vturb = 18.2$ km s$^{-1}$ as the mean $\vturbTh$ found above may thus be a reasonable value to use for application of our model in future works.
Similarly, a reasonable range of variations on this would be 14.8~km~s$^{-1} < \vturb <$ 21.6 km s$^{-1}$ as suggested by plus or minus one standard deviation about this mean.
While this will not provide a perfect match to the mass-loss rates of individual objects (see for example the right panel of Figure \ref{fig:vturb_Obs_theory}) it provides a representative prediction for the average mass-loss behavior of red supergiants such that it would be appropriate to use in stellar evolution codes.
As such, and as discussed further in the following section, future work should be dedicated to improving this estimation, establishing whether $\vturb$ varies in time for individual stars, and testing its dependence on key stellar parameters.

%\section{Recapitulation of results and discussion of future directions}\label{sec:outlook}
\section{Summary and future directions}\label{sec:outlook}
%\begin{itemize}
%    \item Calibrate $v_{turb}$
%    \item $v_{turb}$ inferred supersonic and used supersonic in the MARCS etc models, what does this actually mean for the mass loss structure?
%    \item Dust sublimation radius
%\end{itemize}

We have here derived a theoretical, analytic (dust-free) model for the mass loss-rates of red supergiants. 
Building upon the works by \cite{GusPle92} and \cite{JosPle07}, we examine the role of turbulent pressure in levitating material in the stellar atmosphere up to a modified equivalent of the standard Parker wind critical point.
This is achieved by employing a standard decomposition of pressure into thermal and turbulent pressure components as, for instance, employed in standard 1D spherically symmetric model atmospheres \citep[e.g.][]{GusEdv08}.
By applying the resulting expression for pressure in the steady-state equations for conservation of mass and momentum, we examine both hydrostatic and low velocity levitation for an isothermal model (Sect. \ref{sec:mdot_iso}).
We also numerically solve the full 1D steady-state equation-of-motion for a non-isothermal model employing the \cite{Luc71} temperature structure as a stand in for an energy equation.
By combining each of these cases of steady-state levitation with the constraint that the total accumulated optical depth of circumstellar material from infinity down to the stellar photosphere at $R_\ast = r (\tau = 2/3)$, we provide unique mass-loss rates.

As shown in Sect. \ref{sec:mdot_iso} this method provides a fully analytic mass-loss rate for an isothermal wind.
%Further, this same set of criteria can be iterated over to numerically converge on mass-loss rates in the non-isothermal case, as shown by \S\ref{sec:mdot_noniso}.
%\jon{Building upon the works by Gustafsson et al. (1992) and Josseline \& Plez.. , this model ...  in my opinion, you need to do a few-sentence long executive summary of model here. -- levitation, turbulent driving force, steady-state, etc., and then root it in previous work so that people don't get scared..} 
%By performing a standard decomposition of pressure into thermal and turbulent components, and then applying typical values of the turbulent velocity as inferred from observations we show that red supergiants naturally should develop vigorous mass loss in slow winds in a turbulence driven equivalent of the thermal pressure driven Parker wind.
%As is the case with the Parker wind, 
%As shown in \S\ref{sec:mdot_iso} and \S\ref{sec:mdot_noniso}, the mass-loss rate from red supergiants is therefore fully analytic under the assumption of an isothermal outflow, and can be readily solved for non-isothermal outflows numerically.
%We have therefore performed this latter step for 
Using a grid of numerical models, we %a grid of models to provide a 
further provide a fit to the correction factor between the isothermal and non-isothermal mass-loss rate for the same input parameters (Sect. \ref{sec:noniso_correction}).
Combining this correction factor with the isothermal mass-loss rates then presents a fully analytic mass-loss rate prescription for red supergiants,
%This predictive mass-loss rate is 
given by the combination of Equations \ref{eqn:Rpmod}, \ref{eqn:mdot_an}, \ref{eqn:rhoRp}, and \ref{eqn:corr_fit_2}.
Comparisons of this mass-loss rate with both observations and empirical mass-loss rate prescriptions show that such a turbulent pressure driven mass-loss rate is able to recover realistic mass-loss rates over the range of parameters appropriate to red supergiants.
These comparisons also inform our suggested choice of $\vturb = 18.2 \pm 3.4$ km s$^{-1}$ as an appropriate representative value of $\vturb$ in mass-loss rate calculations where this parameter is not well constrained.

To continue the discussion, we enumerate some items of future work required to place this analytic, theoretical model of red supergiant mass-loss on even firmer footing.
From an observational side, as an immediate first item estimations of turbulent velocities for a wider body of red supergiants are needed.
As it stands, the limited sample that we have examined here provides highly encouraging results, and additional objects would allow the model to be calibrated more carefully.
Additionally, such observations would be of key importance to understand how this current free parameter of the model scales with stellar parameters.
Finally, such future observations could also help constrain the depth dependence of $\vturb$.
Prior work has suggested that turbulent velocity may actually increase with distance from the star over at least part of the wind launching region \citep{deKdeJ88,JosPle07}, and better understanding this profile would be highly useful in further developing the model presented here.
%Especially important would be efforts to infer a depth dependence to $\vturb$ as this would be highly useful for the further development of the model presented here.

To further develop the model itself, main efforts may focus on relaxing key approximations made in this paper.
One of the first of these would be to use a radial profile of $\vturb$, perhaps as inferred by observations above, rather than the spatial average employed here.
Such a radial profile might arise from dissipation of turbulence by shocks.
Inclusion of a radial profile in $\vturb$ would allow, for instance, for a prediction of the terminal velocities of red supergiant winds instead of only mass-loss rates, as the energy available to drive the wind would no longer diverge with increasing outer radius of the model.
Further generalization of $\vturb$ could also include variations over the surface of the star and/or in time.
Both of these would arise naturally if turbulence is related to surface convection in the star, and their inclusion would help to explain the observed clumpy structure of the wind.

%Such a radial profile would also be important for a second important line of enquiry, namely an investigation of the implications for the overall wind structure of driving with supersonic turbulence.
%Such supersonic turbulent velocities are commonly employed in model atmospheres and are inferred by observations, and it is possible that breaking of these supersonic motions would help to explain the observed clumpy mass loss from red supergiants as local regions around the star obtain enhanced and damped turbulent pressure contributions.

A further investigation of the overall structure of red supergiant winds may also re-introduce dust and/or molecules and their associated opacity to examine the impact this would have for the proposed model.
As discussed in the introduction, previous models have assumed that radiation pressure is the dominant driving force in carrying the mass loss of red supergiants, with the main question being the origin of the ``missing force'' necessary to get material from the stellar photosphere up to the region where molecules and dust can form, the latter occurring potentially several stellar radii away from the photosphere.
Seemingly circumventing such a model, we show that turbulent pressure extended red supergiant atmospheres are already able to generate the appropriate mass-loss rates in the absence of radiation pressure on dust or molecules.
However, it is important to recognize that this does not necessarily mean that enhanced opacity from dust or molecules has a negligible contribution.
In fact, these opacities almost certainly play a crucial role in setting the wind terminal speed as turbulent pressure dies off away from the stellar surface (see the discussion in Sect. \ref{sec:noniso_iter}).
It is also clear, as discussed in Sect. \ref{sec:iso_scaling}, that the overall scale of opacity and its radial profile may contribute to the optical depth and scale height of the wind and thus to the mass loss rate retrieved by this model.
Further, it is even possible that the radial profile of turbulent pressure is such that turbulent pressure plays the dominant role in levitating material off the stellar surface, but that dust or molecular opacity carries the wind through the critical point as previously theorized.

Another important line of investigation regards the origin of the turbulent velocities and extended atmospheres this model leverages.
As discussed by \cite{ArrWit15}, previous models have focused on convection through 3D, so-called ``star-in-a-box'' radiation hydrodynamics simulations such as those computed with CO$^{5}$BOLD \citep{FreSte12}, and on pulsations through 1D simulations akin to those used for AGB stars and especially Mira variabls \citep[e.g. \code{CODEX}][]{IreSch08,IreSch11}. 
While convection and pulsations in the stellar atmosphere both provide tempting options, convection simulations generally fail to generate the turbulent velocities or atmospheric extensions inferred from observations, and pulsations are required to be of unrealistically high amplitude compared to observed lightcurves for red supergiants \citep[see, e.g.][]{JosPle07,ArrWit15}.

Similarly, one can examine the role that rotation might have in generating similar levels of extension to the stellar atmosphere.
This can be done by applying the centrifugal term ($\Omega^2 r$) either along side or instead of $\vturb$ in the preceding analysis.
Taking the limiting case where turbulent pressure is omitted, we find that matching the atmospheric extension of a 18.2 $\rm km s^{-1}$ turbulent velocity even at the most optimal position on the stellar equator (including the effects of stellar oblateness) requires a rotation velocity almost $94\%$ of the orbital velocity, implausibly fast for a red supergiant.
As we reinforce through this paper that atmospheric extension is crucial for the mass-loss process, further examination of the possible missing ingredients for such models as well as potential constructive interactions between them is now highly timely.

To wrap up the discussion, we address some of the potential implications of applying our mass-loss rates, especially for the case of stellar evolution modeling.
As discussed in Sect. \ref{sec:comp_w_emp_rates}, the current state-of-the-art mass-loss rates for stellar evolution modeling are all empirical.
Given that none of the empirical rates match the theoretical one we present here for all stars, these new theoretical rates should impact the mass, luminosity, and radius distributions of supergiant supernova progenitors.
Additionally, the steeper trend of increasing mass-loss rate with increasing luminosity would naturally turn more massive stars evolving toward the red supergiant branch back to the blue due to envelope stripping, thereby predicting a decreased upper luminosity limit for red supergiants.
Finally, these rates provide intriguing implications for stellar evolution in lower metallicity environments.
Many standard stellar evolution implementations of mass loss in the red supergiant branch impose a downturn with metallicity partially inspired by the $\dot{M} \propto Z^n$ dependence of line-driven hot star winds (\citealt{VindeK01,MokdeK07,BjoSun20}).
Observations analyzed by \cite{MauJos11} do seem to suggest such a scaling of $\dot{M} \propto \left(Z/Z_\odot\right)^{0.7}$ from the Galaxy to the red supergiants in the Small Magellanic Cloud.
However, they concluded that the scaling to the Large Magellanic Cloud was not well constrained by the samples they considered, emphasizing that more investigation is needed.
Meanwhile, the model we present here should not be strongly dependent on metallicity.
While the optical depth and scale height of the wind clearly do depend on dust and molecular content, which in turn depend on metallicity, this is a weaker secondary scaling when compared with the turbulent pressure itself.
Moreover, decreasing opacity can lead to increased mass-loss rates in this model depending where in parameter space a star sits (see Sect. \ref{sec:iso_scaling}).
So long as the turbulent pressure mechanism itself is not strongly metallicity dependant, which turbulence triggered by hydrogen recombination is not, this could then imply much stronger mass loss on the red supergiant branch at low metallicity than previously assumed, and as such significantly different stellar evolution tracks.

\begin{acknowledgements}
We would like to thank the anonymous referee for their comments which enhanced the discussion in this paper.
We acknowledge support from the KU Leuven C1 grant MAESTRO C16/17/007. L.D. also acknowledges funding from the European Research Council (ERC) under the European Union's Horizon 2020 research and innovation programme (grant agreements No.\ 646758: AEROSOL with PI L.\ Decin).
\end{acknowledgements}

\bibliographystyle{aa}
\bibliography{biblio}

\appendix
\section{Computation of $\gRad$ and $\kappa$ from accelerating molecules}\label{app:line_force}

%\dylan{
As discussed in Sect. \ref{sec:mdot_iso}, the total opacity scale of the winds we treat here can be strongly impacted by the Doppler shifting of spectral lines, leading to opacities exceeding the static Rosseland mean opacity.
To show this, we begin with a generalized form of radiative acceleration in the $\hat{n}$ direction,
%}

\begin{equation}
    \mathbf{g_\mathrm{rad}} = \frac{1}{c} \oint \int_0^\infty \kappa_\nu\, I_\nu\, d \nu\, \mathbf{\hat{n}}\, d\Omega\;,
\end{equation}
%\dylan{
with speed of light $c$, extinction coefficient $\kappa_\nu$ [$\rm cm^2\; g^{-1}$], 
and intensity per unit frequency $I_\nu$.
If one assumes that the extinction comes from a single, isolated spectral line, then a natural choice is to split $\kappa_\nu$ into a normalized, frequency-dependent shape of the spectral line, or profile function $\phi_\nu$, and a line-integrated total extinction coefficient 
%}
\begin{equation}
    \kappa_L = \frac{\sigma_{cl}\, n_l\, f_{lu}}{\rho}\left(1 - \exp\left(-\frac{h\, \nu_\mathrm{o}}{k_B\, T}\right)\right)\;.
\end{equation}
%\dylan{
In this expression for $\kappa_L$, $\sigma_{cl}$ is the classical oscillator cross section, $n_l$ is the number density of particles in the lower level of the transition, $f_{lu}$ is the quantum mechanical oscillator strength of the transition, and $\rho$ is mass density.
The final exponential term accounts for stimulated emission under the assumption of local thermodynamic equilibrium (LTE). Such stimulated emission can play 
%which plays 
an important role for the considered infrared spectral lines transitions as the energy of the transition, given by Planck's constant $h$ times rest frequency $\nu_\mathrm{o}$, can be comparable to the thermal energy of the gas, given by Boltzmann's constant $k_B$ times gas temperature $T$.
Note that here $\kappa_L$ has picked up units of [$\rm cm^2\; g^{-1}\; Hz$] as the profile function is defined per unit frequency.
Under these substitutions, the equation we now need to solve becomes
%}

\begin{equation}
    \mathbf{g_\mathrm{line}} = \frac{\kappa_L}{c} \oint \int_0^\infty \phi_\nu\, I_\nu\, d \nu\, \mathbf{\hat{n}}\, d\Omega\;.
\end{equation}

%\dylan{
At this point, it is important to recognize that the intensity fed into this expression can itself be a strong function of frequency even in the presence of an optically thin continuum, as is the case for red supergiants, due to the attenuation of intensity by the spectral line we consider itself.
Therefore, we have to obtain intensity from the formal solution of the radiative transfer equation
%}

\begin{equation}
    I_\nu = I_\nu^\mathrm{o} e^{-\tau_\nu} + \int S_\nu(t_\nu)\, e^{-\lvert \tau_\nu - t_\nu \rvert}\, dt_\nu\;.
\end{equation}
%\dylan{
While in general this becomes quite complex as the source function $S_\nu$ depends on $I_\nu$, we can simplify by leveraging symmetry arguments.
In the Sobolev limit considered here (see further below), the second term here becomes fore-aft symmetric such that it cancels out in the integral $\oint \hat{n} d\Omega$.
As regards the optical depth $\tau_\nu$, we take the approximation that this arises only from extinction from the spectral line itself such that
%}

\begin{equation}
    \tau_\nu = \int \kappa_L\, \phi_\nu\, \rho\, dl\;,
\end{equation}
%\dylan{
where the mixed frequency and spatial notations can be rectified by using the Doppler formula
%}

\begin{equation}
    \frac{\nu - \nu_\mathrm{o}}{\nu_\mathrm{o}} = \frac{v_l}{c}\;,
\end{equation}
%\dylan{
such that
%}

\begin{equation}
    \tau_\nu = \int \kappa_L\, \phi_\nu\, \rho \left(\frac{\partial l}{\partial v_l}\right)\left(\frac{\partial v_l}{\partial\nu}\right) d\nu = \int \frac{\kappa_L\, \rho\, c}{\nu_\mathrm{o}\, \partial v_l/\partial l}\, \phi_\nu\, d\nu\;.
\end{equation}
%\dylan{
Finally, we can take the Sobolev approximation \citep{Sob60} which argues that all hydrodynamic variables are effectively constant over a line resonance region and therefore can be pulled through the frequency integral to give
%}

\begin{equation}
    \mathbf{g_\mathrm{line}} = \frac{\kappa_L}{c} \oint \int_0^\infty \phi_\nu\, I_\nu^\mathrm{o}\, e^{-\tau_S \int_0^\infty \phi_\nu\, d\nu} d \nu\, \mathbf{\hat{n}}\, d\Omega\;,
\end{equation}
%\dylan{
where we have introduced the Sobolev optical depth $\tau_S~=~\kappa_L \rho c/(\nu_\mathrm{o} \partial v_l/\partial l)$. 
This final expression can now be analytically solved to give
%}

\begin{equation}
    \mathbf{g_\mathrm{line}} = \frac{\kappa_L}{c}\oint I_\nu^\mathrm{o} \left(\frac{1-e^{-\tau_S}}{\tau_S}\right) d\Omega\;,
\end{equation}
%\dylan{
which further simplifies in the case of a spherically symmetric flow and radiation from a point star to
%}

\begin{equation}
    g_\mathrm{line,r} = \frac{\pi\, \kappa_L\, I_\nu^\mathrm{o}}{c} \left(\frac{1-e^{-\tau_{S,r}}}{\tau_{S,r}}\right)\;,
\end{equation}
%\dylan{
with $\tau_{S,r}$ replacing the general line of sight velocity gradient $\partial v_l/\partial l$ in $\tau_S$ with the radial velocity gradient $\partial v_r/\partial r$.

The Sobolev approximation which we have employed here is valid in wind outflow regions where the flow speed exceeds a few times the characteristic thermal speed, $v_{\rm th}$, such that there is not significant absorption of intensity entering the resonance region by a spectral line in the hydrostatic stellar atmosphere, and such that the physical size of the resonance region is small compared to the scale over which the hydrodynamic quantities entering $\tau_S$ vary.
In practice, this condition is nearly always met, however, as the thermal speeds of the molecules we examine are reduced compared to the bulk wind sound speed by the potentially quite substantial ratio of mean molecular mass of the bulk gas to mass of the molecule.
Therefore, as this expression holds for each individual line over the bulk of the wind, to generalize this to a spectrum of lines one simply needs to sum $g_\mathrm{line}$.
Numerous tabulated line lists are available to facilitate this process, each including $\nu_\mathrm{o}$, either $f_{lu}$ or the Einstein coefficients necessary to compute it, and the energy and degeneracy of the levels involved in each of the transitions, which can be used to compute $n_l$, for instance from the Boltzmann distribution in Local Thermodynamic Equilibrium.
Taking this fully summed acceleration, it is then straightforward to back out a flux-weighted mean opacity $\kappa$ of the line acceleration by appealing to Equation \ref{eqn:g_rad} such that
%}

\begin{equation}
    \kappa = \frac{c}{F_\mathrm{r}} \sum g_\mathrm{line,r}\;,
\end{equation}
where the summation is over all lines.
As more force is available from an accelerating spectral line than a static one, this will always return flux-weighted mean opacity that is larger than the Rosseland mean opacity, thereby substantiating our choice to test the effects of varying $\kappa$ in Sect.~\ref{sec:iso_scaling}.

\end{document}